\documentclass[a4paper,11pt]{article}

\usepackage{jcappub}
\usepackage{soul}

\newcommand{\bk}{{\mathbf k}}

\newcommand{\bn}{{\mathbf n}}

\newcommand{\bV}{{\mathbf V}}

\newcommand{\MM}{{\cal M}}

\newcommand{\HH}{{\cal H}}

\newcommand{\al}{\alpha}

\newcommand{\de}{\delta}
\newcommand{\De}{\Delta}

\newcommand{\ga}{\gamma}

\newcommand{\ka}{\kappa}

\newcommand{\La}{\Lambda}
\newcommand{\la}{\lambda}
\newcommand{\Om}{\Omega}
\newcommand{\om}{\omega}
\newcommand{\si}{\sigma}

\newcommand{\ra}{\rightarrow}

\newcommand{\be}{\begin{equation}}
\newcommand{\ee}{\end{equation}}

\newcommand{\bea}{\begin{eqnarray}}
\newcommand{\eea}{\end{eqnarray}}
\newcommand{\bean}{\begin{eqnarray*}}
\newcommand{\eean}{\end{eqnarray*}}
\newcommand{\dd}{\partial}

\newcommand*\bra[1]{\left(#1 \right)}
\newcommand*\sbra[1]{\left[#1 \right]}
\newcommand*\bbra[1]{\left\{#1 \right\}}

\newcommand{\class}{{\sc class}}
\newcommand{\classgal}{{\sc class}gal}

\usepackage[usenames,dvipsnames]{xcolor}

\usepackage{ulem}

\title{Cosmological Parameter Estimation with Large Scale Structure Observations}

\author[a]{Enea Di Dio, }
\author[a]{Francesco Montanari, }
\author[a]{Ruth Durrer, }
\author[b,c,d]{Julien Lesgourgues, }

\affiliation[a]{D\'epartement de Physique Th\'eorique and Center for Astroparticle Physics,
Universit\'e de Gen\`eve\\ 24 quai Ernest  Ansermet, 1211 Gen\`eve 4, Switzerland}
\affiliation[b]{Institut de Th\'eorie des Ph\'enom\`enes Physiques, \'Ecole Polytechnique F\'ed\'erale de Lausanne,\\ CH-1015, Lausanne, Switzerland}
\affiliation[c]{CERN, Theory Division,\\ CH-1211 Geneva 23, Switzerland}
\affiliation[d]{LAPTh (CNRS -Universit\'e de Savoie),\\ BP 110, F-74941 Annecy-le-Vieux Cedex, France}

\emailAdd{Enea.DiDio@unige.ch}
\emailAdd{Francesco.Montanari@unige.ch}
\emailAdd{Ruth.Durrer@unige.ch}
\emailAdd{Julien.Lesgourgues@cern.ch}

\abstract{
We estimate the sensitivity of future galaxy surveys to cosmological parameters, using the redshift dependent angular power spectra of galaxy number counts, $C_\ell(z_1,z_2)$, calculated with all relativistic corrections at first order in perturbation theory.
We pay special attention to the redshift dependence of the non-linearity scale and present Fisher matrix forecasts for Euclid-like and DES-like galaxy surveys.  We compare the standard $P(k)$ analysis with the new $C_\ell(z_1,z_2)$ method. We show that for surveys with photometric redshifts the new analysis performs significantly better than the $P(k)$ analysis. For spectroscopic redshifts, however, the
large number of redshift bins  which would be
needed to fully profit from the redshift information, is severely limited by  shot noise.
We also identify surveys which can measure the lensing contribution and we study the monopole, $C_0(z_1,z_2)$.
}

\keywords{Cosmology: theory, forecasts, Large Scale Structure
\vskip13pt plus8pt minus11pt
\noindent{\bfseries\large\sffamily{Preprints:}} CERN-PH-TH/2013-155, LAPTH-037/13}

\begin{document}
\maketitle
\flushbottom

\section{Introduction}\label{s:intro}
Observations and analysis of the  cosmic microwave background (CMB) have led to stunning advances in observational cosmology~\cite{Durrer:2008aa,Ade:2013zuv}.  This is due on the one hand to an observational effort which has led to excellent data, but also to the theoretical simplicity of CMB physics. In a next (long term) step, cosmologists
will try to repeat the CMB success story with observations of large scale structures (LSS), i.e. the distribution of galaxies in the Universe.

The advantage of LSS data is the fact that it is three-dimensional, and therefore contains much more information than the two-dimensional CMB. The disadvantage is that the interpretation of the galaxy distribution is much more complicated than that of CMB anisotropies. First of all, our theoretical cosmological models predict the fluctuations of a continuous density field, which we have to relate to the discrete galaxy distribution. Furthermore, on scales smaller than $30h^{-1}$Mpc,  matter density fluctuations become large and linear perturbation theory is not sufficient to compute them. On these scales, in principle, we rely on costly N-body simulations.

In an accompanying paper~\cite{Didio1} we describe a code, CLASSgal, which calculates galaxy number counts,
$\De(\bn,z)$, as functions of direction $\bn$ and observed redshift $z$ in linear perturbation theory. In this code all the relativistic effects due to peculiar motion, lensing, integrated Sachs Wolfe effect (ISW) and other effects of metric perturbations as described in~\cite{Bonvin:2011bg,Challinor:2011bk} are fully taken into account.
Even if a realistic treatment of the problem of biasing mentioned above is still missing,  the number counts have the advantage that they are directly observable as opposed to the  power spectrum of fluctuations in real space which depends on cosmological parameters. The problem how the galaxy distribution, number counts and distance measurements are affected by
the propagation of light in a perturbed geometry has also been investigated in other works; see, e.g.~\cite{Yoo:2009au,Yoo:2010ni,Jeong:2011as,Schmidt:2012ne,Bertacca:2012tp,Jeong:2013psa}.

In this paper we use CLASSgal to make forecasts for the ability to measure cosmological parameters from Euclid-like and DES-like galaxy surveys. This also helps to determine optimal observational specifications for such a survey.
The main goal of the paper is to compare the traditional $P(k)$ analysis of large scale structure with the new $C_\ell(z_1,z_2)$ method.
To do this we shall study and compare the figure of merit (FoM) for selected pairs of parameters. As our goal is not a determination of the cosmological parameters but a comparison of methods, we shall not use constraints on the parameters from the Planck results or other surveys. We just use the Planck best fit values as the fiducial values for our Fisher matrix analysis. We mainly want to analyze the sensitivity of the results to redshift binning, and to the inclusion of cross correlations, i.e., correlations at different redshifts. We shall also study the signal-to-noise of the different contributions to $C_\ell(z_1,z_2)$ in order to decide whether they are measurable with future surveys.

In the next section we exemplify how the same number counts lead to different 3D power spectra when different cosmological parameters are employed. In Section~\ref{s:fis} we use the Fisher matrix technique to estimate cosmological parameters from the number count spectrum. We pay special attention on the non-linearity scale which enters in a non-trivial way in the Fisher matrix. We then determine the precision with which we can estimate cosmological parameters for different choices of redshift bins and angular resolution. In Section~\ref{s:res} we discuss our results and put them into perspective.
In Section~\ref{s:con} we conclude.
Some details on the Fisher matrix technique and the basic formula for the number counts are given in the appendix.

\section{Number counts versus the real space power spectrum}
\begin{figure}[tbp]
\includegraphics[width=0.5\linewidth]{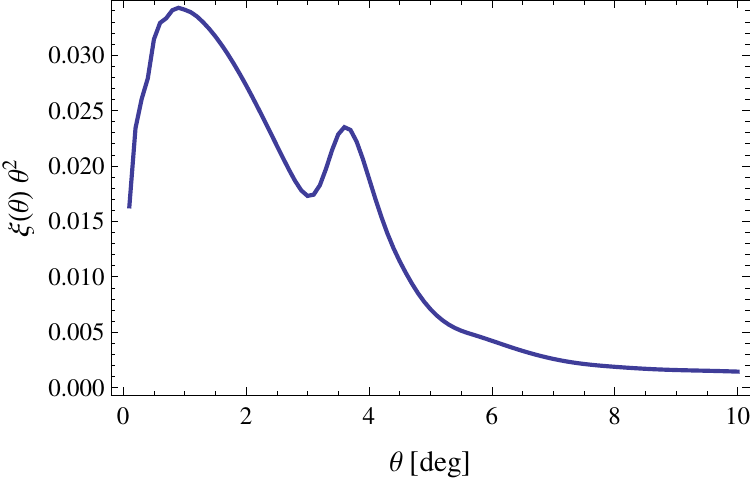}~
\includegraphics[width=0.5\linewidth]{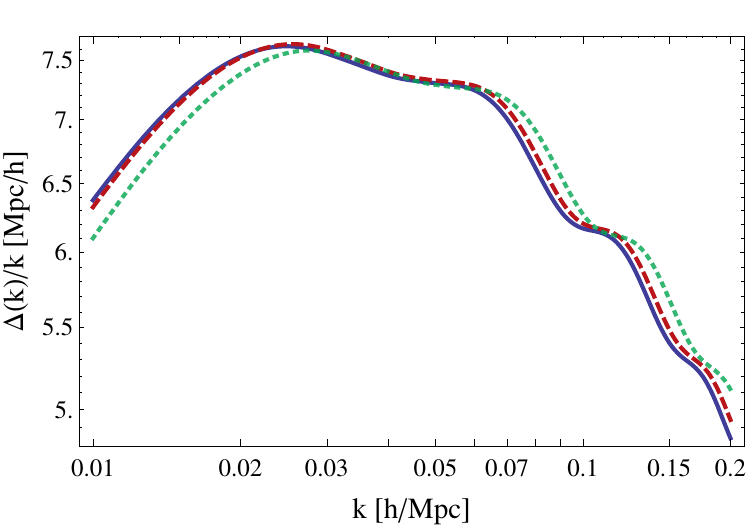}
\caption{\label{f:Psex} Left: the observable correlation function multiplied by $\theta^2$, calculated according to \cite{Montanari:2012me} for $\Om_m=0.24$ and $z_1=z_2=0.7$.
Right: the power spectrum $\De^2(k)=k^3P(k)/2\pi^2$ divided by $k$ and scaled to $z=0$, obtained by Fourier transforming the correlation function on the top, using three different assumptions to convert $\bbra{\theta,z_1,z_2}$ into comoving distances: either the fiducial value $\Om_m=0.24$ (solid, blue line), or two different values $\Om_m=0.3,0.5$ (dashed, red line and dotted, green line, respectively). All other cosmological parameters are fixed to the WMAP7 best fit values.
}
\end{figure}

In this section we illustrate explicitly  that the 3D power spectrum is not directly observable. This fact is not new, but it seems not to prevent the community from using the power spectrum
which itself depends on cosmological parameters to estimate the latter.  This is then usually taken into account by a recursive method: one chooses a set of cosmological parameters (usually the best fit parameters from CMB observations), determines the power spectrum under the assumption that this set correctly describes the background cosmology, and then estimates a new set of cosmological parameters. This process is repeated with the new parameters until convergence is
reached, see~\cite{Blake:2011ep,Reid:2012sw,Nuza:2012mw,Anderson:2013oza} and others. It is certainly possible to find the best fit parameters in this way, but a correct determination of the errors is more complicated, since not only the power spectrum but also its argument $k$ depends on cosmological parameters.
In Fig.~\ref{f:Psex} we show what happens when a measured correlation function is converted into a power spectrum using the wrong cosmological parameters.

The advantage of using the power spectrum is that different Fourier modes are in principle independent. Therefore, errors on the power spectrum are independent. Furthermore,  for small redshifts, $z\ll 1$, the parameter dependence of the distance is simply $r(z) \simeq H_0^{-1}z,~ z\ll 1$, where $H_0$ denotes the Hubble constant. This is usually absorbed by measuring distances in units of $h^{-1}$Mpc. However,  for large redshifts the relation becomes more complicated,
\bea
r(z) &=& \int_0^z\frac{dz'}{H(z')}\,, \eea
where $H(z)= H_0\left(\Om_m(1+z)^3 + \Om_K(1+z)^2 +\Om_{\rm DE}(z)\right)^{1/2}$,
 $\Om_m=8\pi G\rho_m/(3H_0^2)$ denotes the present matter density parameter,
 $\Om_K = -K/H_0^2$ is the present curvature parameter, and $\Om_{\rm DE}(z)=8\pi G\rho_{\rm DE}(z)/(3H_0^2)$ is the dark energy density parameter. $\rho_{\rm DE}$ is in principle redshift dependent, but for a cosmological constant we simply have
$$ \rho_{\rm DE} = \rho_\La = \frac{\La}{8\pi G}\,,\qquad  \Om_{\rm DE}=\frac{\La}{3H_0^2}.$$
Furthermore, we see perturbations which are far away from us, at an earlier stage of their evolution, when the power spectrum had a different amplitude.

We therefore propose here an alternative method to analyze large scale structure data.
What we truly measure in a galaxy survey is the position of each galaxy in the sky, given by a direction $\bn$ and a redshift $z$.

If we consider only the two dominant contributions, namely the density fluctuations and the redshift space distortions~\cite{Kaiser1987,Hamilton1992,Raccanelli:2010hk} in the number counts, we have~\cite{Bonvin:2011bg,Montanari:2012me}
\be\label{e:Dereal}
\De(\bn,z) = D(r(z)\bn, t(z)) + \frac{1}{\HH(z)}\dd_r\left(\bV(r(z)\bn, t(z))\cdot\bn\right) \,.
\ee
Here $D$ is the density fluctuation in comoving gauge and $\bV$ is  the peculiar velocity  in longitudinal gauge. The full expression for $\De(\bn,z)$ including lensing and other subdominant contributions from peculiar velocity and the gravitational potential is given in Appendix~\ref{appA}.
Both $r(z)$ and the conformal time $t(z) \equiv t_0-r(z)$ depend on the cosmology. To first order in perturbation theory, we can neglect the fact that the redshift and position ($z, \bn)$ that appear in the argument of  $D$ and $\bV$ are also perturbed.

Fourier transforming Eq.~(\ref{e:Dereal}), and noting that in Fourier space $\dd_r \ra i\bk\cdot\bn$, we obtain
\be\label{e:Defour}
\De(\bk,t(z)) = D(\bk, t(z)) + \frac{k}{\HH(z)}(\hat\bk\cdot\bn)^2V(\bk, t(z)) \,.
\ee
Here $V$ is the velocity potential such that $\bV(\bk) =i\hat\bk V$ and $\mu =(\hat\bk\cdot\bn) =\cos\phi$, where $\phi$ denotes the angle between $\bk$ and $\bn$ and $\hat\bk =\bk/k$ is the unit vector in direction $\bk$. For pure matter perturbations, the time dependence of  $D$ is independent of wave number such that
$$ D(\bk, t(z)) = G(z)D(\bk,t_0)\,,$$
where $G(z)$ describes the growth of linear matter perturbations. For pure matter perturbations, the continuity equation~\cite{Durrer:2008aa}   implies
 \be
 \frac{k}{\HH(z)}V(\bk, t(z)) = f(z)D(\bk, t(z)) \,,
~\text{ where }~
  f(z) = \frac{1+z}{G}\frac{dG}{dz} = \frac{d\log G}{d\log(1+z)}
\ee
is the growth factor. Inserting this in Eq.~(\ref{e:Defour}), we obtain the following relation for a fixed angle between $\bk$ and the direction of observation, and for a fixed redshift $z$:
\be\label{e:Pfour}
P_\De(k,z) = G^2(z)\left(1 +\mu^2f(z)\right)^2P_D(k) \,.
\ee
If we have only one redshift bin at our disposition, the factor $G^2$ is in principle degenerate with a constant bias and the overall amplitude of $P_D(k)$. A more general expression for different redshifts and directions can be found in Ref.~\cite{Montanari:2012me}.

Measuring the redshift space distortions that are responsible for the angular dependence of $P_\De$ allows in principle to measure the growth factor, $f(z)$.
Furthermore, assuming that density fluctuations relate to the galaxy density by some bias factor $b(z,k)$, while the velocities are not biased, the first term in Eq.~\ref{e:Pfour} becomes proportional to $(b(z,k)G(z))^2$ while the second term behaves like $G(z)f(z)$.
This feature allows in principle to reconstruct the bias function $b(z,k)$. It is an interesting question whether this bias can be measured better with our angular analysis or with the standard power spectrum analysis. However, since it is probably not strongly dependent on cosmological parameters, its reconstruction with the power spectrum method seems quite adequate and probably simpler. In the remainder of the present work we shall therefore not address the interesting question of bias, see~e.g.~\cite{Carbone:2010sb,Desjacques:2012eb}.

However, in the above approximation, both relativistic effects which can be relevant on large scales as well as a possible clustering of dark energy are neglected. On top of that, the matter power spectrum in Fourier space is not directly observable. Several steps (like the relation between Fourier wave numbers and galaxy positions) can only be performed under the assumption of a given background cosmology.
What we truly measure is a correlation function $\xi(\theta,z,z')$, for galaxies at redshifts $z$ and $z'$ in directions $\bn$ and $\bn'$, where  $\theta$ is the angle between the two directions, $\bn\cdot\bn'=\cos\theta$.
The  correlation function analysis of cosmological surveys has a long tradition. But the correlation function has usually been considered as a function of the distance $r$ between galaxies which of course has the same problem as its Fourier transform, the power spectrum: it depends on the cosmological parameters used to determine $r$.

For these reasons, we work instead directly with the correlation function and power spectra in angular and redshift space. They are related by
\bea
\xi(\theta,z,z')  \equiv  \langle \De(\bn,z) \De(\bn',z')\rangle
&=& \frac{1}{4\pi}\sum_{\ell=0}^{\infty} (2\ell+1)C_\ell (z,z')P_\ell(\cos\theta) \,, \label{e:corfun}
\eea
where $P_\ell(x)$ is the Legendre polynomial of degree $\ell$.
The power spectra $C_\ell (z,z')$ can also be defined via
\be
\De(\bn,z) =\sum_{\ell m}a_{\ell m}(z)Y_{\ell m}(\bn)  \,,
\qquad
 C_\ell (z,z') = \langle a_{\ell m}(z) a^*_{\ell m}(z')\rangle  \,,
\ee
where  the star indicates complex conjugation.

The full expression relating $\De(\bn,z)$ or $ C_\ell (z,z')$ to the the primordial power spectrum (given e.g.~by inflation), valid at first order in perturbation theory and taking into account all relativistic effects,  can be found in Refs.~\cite{Didio1} or~\cite{Bonvin:2011bg}.

The disadvantage of this quantity w.r.t the power spectrum is the fact that the $C_\ell$'s at different redshifts are not independent. Their correlation is actually very important as it encodes, e.g., the radial BAO's (Baryon Acoustic Oscillations).

We finish this section with the conclusion that in order to measure quantities which are virtually independent of cosmological parameters the power spectrum analysis is sufficient and probably simpler, however when we want to constrain cosmological parameters or related quantities like the growth factor, the angular method proposed here is safer.
It has the advantage that it is fully model independent.

\section{The Fisher matrix and the nonlinearity scale}\label{s:fis}
We consider galaxy number counts as functions of the observational direction and the observed redshift, $\De(\bn,z)$. In Ref.~\cite{Didio1} we describe how the code CLASSgal  calculates the corresponding power spectrum $C_\ell(z_1,z_2)$.
We shall consider these spectra as our basic observables and assume that different $\ell$-values are uncorrelated.
The truly observed spectra have finite resolution in redshift, and are of the form
\be
C^W_\ell \! (z,z') = \int W(z_1,z,\De z)  W(z_2,z'\!,\De z)C_\ell (z_1,z_2)dz_1dz_2\,.
\ee
Here $W(z_1,z,\De z)$ is a normalized window function centered around $z$ with half-width $\De z$. We shall use Gaussian and top hat windows (with half-width $\De z$). Here
$C^W_\ell$ are the Legendre coefficients of the smoothed angular correlation function
 \be
\xi^W(\theta,z,z') =  \int W(z_1,z,\De z)  W(z_2,z'\!,\De z) \xi(\theta,z_1,z_2)dz_1dz_2\,.
 \ee

\subsection{ Fisher matrix forecasts}
We perform a Fisher matrix analysis to compare a forecast for future redshift surveys derived from the angular power spectrum $C_\ell (z_1 ,z_2)$ with the one derived from
the three-dimensional Fourier power spectrum $P(k)$. For a given list of $N_{\text{bin}}$ redshift bins with mean redshifts $z_i$, we denote the auto- and cross-correlation angular power spectra by $C_\ell^{ij}\equiv C^W_\ell(z_i,z_j)$.  Since $C_\ell^{ij}=C_\ell^{ji}$, there are $[N_{\text{bin}}(N_{\text{bin}}+1)/2]$ power spectra to be considered. Assuming that the fluctuations are statistically homogeneous, isotropic and Gaussian distributed, the covariance matrix between different power spectra can be approximated as in Ref.~\cite{Asorey+12},
\be\label{e:cov}
\text{Cov}_{[\ell,\ell'] [(ij), (pq)]}=\de_{\ell,\ell'}\frac{C_\ell^{\text{obs},i p} C_\ell^{\text{obs},jq} + C_\ell^{\text{obs},i q} C_\ell^{\text{obs},jp}}{f_\text{sky} \left( 2 \ell + 1 \right) },
\ee
where $f_\text{sky}$ is the sky fraction covered by the survey.  For each multipole $\ell$, the covariance matrix is a symmetric matrix,  $\text{Cov}_{[\ell,\ell] [(ij), (pq)]}=\text{Cov}_{[\ell,\ell] [(pq),(ij)]}$, of dimension $[N_{\text{bin}}(N_{\text{bin}}+1)/2)]^2$. The definition of the truly observable power spectrum $C_\ell^{\text{obs} ,ij}$ takes into account the fact that we observe  a finite number  of galaxies instead of a smooth field. This leads to a shot noise contribution in the auto-correlation spectra,
\be
C_\ell^{\text{obs} ,ij}= C_\ell^{ij} + \frac{\delta_{ij}}{\De N(i)},
\ee
where $\De N(i)$ is the number of galaxies per steradian in the $i$-th redshift bin.
In principle also instrumental noise has to be added, but we neglect it here, assuming that it is smaller than the shot noise. More details about the Fisher matrix and the definition of 'figure of merit' (FoM) are given in Appendix~\ref{appB}.

\subsection{The nonlinearity scale}\label{ss:nonlin}

Our code CLASSgal uses linear perturbation theory, which is valid only for small density fluctuations, $D=\de\rho_m/\bar\rho_m\ll 1$. However, on scales roughly of the order of $\la \lesssim 30 h^{-1}$Mpc, the observed density fluctuations are of order unity and larger at late times. In order to compute their evolution, we have to resort to Newtonian N-body simulations, which is beyond the scope of this work. Also, on non-linear scales, the Gaussian approximation used in our expression for likelihoods and Fisher matrices becomes incorrect. Hence, there exists a maximal wavenumber $k_{\max}$ and a minimal comoving wavelength, $$\la_{\min}=\frac{2\pi}{k_{\max}}$$
beyond which we cannot trust our calculations.
There are more involved procedures to deal with non-linearities instead of using a simple cutoff, see e.g.~\cite{Audren:2012vy}. However, in this work we follow the conservative approach of a cutoff.

For a given power spectrum $C_{\ell}^{ij}$  this nonlinear cutoff translates into an $\ell$-dependent and redshift-dependent distance
which can be approximated as follows. The harmonic mode $\ell$ primarily measures fluctuations on an angular scale\footnote{Here we refer to the angular scale separating two consecutive maxima in a harmonic expansion, and not to the scale separating consecutive maxima and minima, given by $\pi/\ell$. Since we want to relate angular scales to Fourier modes, and Fourier wavelengths also refer to the distance between two consecutive maxima, the relation $\theta(\ell)=2\pi/\ell$ is the relevant one in this context.} $\theta(\ell) \sim 2 \pi/\ell$.
Let us consider two bins with mean redshifts $\bar{z}_i\leq\bar{z}_j$ and half-widths $\De z_i$ and $\De z_j$.
At a mean redshift $\bar z = (\bar z_i+\bar z_j)/2$ the scale $\theta(\ell)$ corresponds to a comoving  distance $r_\ell = r(\bar z)\theta(\ell)$, where $r(\bar z)$ denotes the comoving distance to $\bar z$.
Let us define the ``bin separation'' $\de z_{ij}= z_j^{\rm inf}-z_i^{\rm sup}$, where $z_j^{\rm inf}=\bar z_j-\De z_j$ and $z_i^{\rm sup}=\bar z_i+\De z_i$.  Hence $\de z_{ij}$ is positive for well-separated bins, and negative for overlapping bins (or for the case of an auto-correlation spectrum with $i=j$)\footnote{\label{two_or_three} In the case of a spectroscopic survey, we will use top-hat window functions. In this case there is no ambiguity in the definition of $\de z_{ij}$, since ($z_i^{\rm sup}$, $z_j^{\rm inf}$) are given by the sharp edges of the top-hats. In the case of a photometric survey, we will work with Gaussian window functions. Then ($z_i^{\rm sup}$, $z_j^{\rm inf}$) can only be defined as the redshifts standing at an arbitrary number of standard deviations away from the mean redshifts ($\bar{z}_i$, $\bar{z}_j$). In the following, the standard deviation in the $i$-th redshift bin is denoted $\Delta z_i$, and we decided to show results for two different definitions of the ``bin separation'', corresponding either to  $2\si$ distances, with
$z_j^{\rm inf}=\bar z_j-2\De z_j$ and $z_i^{\rm sup}=\bar z_i+2\De z_i$, or to $3\si$ distances, with $z_j^{\rm inf}=\bar z_j-3\De z_j$ and $z_i^{\rm sup}=\bar z_i+3\De z_i$.}.

When $\de z_{ij} < 0$, excluding non-linear scales simply amounts in considering correlations only for distances $r_\ell > \lambda_{\min}$. When  $\de z_{ij} > 0$, the situation is different. The comoving radial distance corresponding to the bin separation $\de z_{ij}$ is $r_z =\de z_{ij} \, H^{-1}(\bar z)$. If $r_z>\lambda_{\min}$, we can consider correlations on arbitrarily small angular scales without ever reaching non-linear wavelengths (in that case, the limiting angular scale is set  by the angular resolution of the experiment). Finally, in the intermediate case such that $0<\de z_{ij} H^{-1}(\bar z) < \lambda_{\min}$, the smallest wavelength probed by a given angular scale is given by $\sqrt{r_\ell^2 + r_z^2}$. In summary, the condition to be fulfilled by a given angular scale $\theta(\ell)$ and by the corresponding multipole $\ell$ is
\bea\label{e:lin}
\la_{\min} \! &\leq&  \! \left\{ \begin{array}{ll}
 \frac{2 \pi}{\ell}r(\bar z)  & \mbox{if}  ~ \de z_{ij} \leq 0  \,,  \\
 & \\
   \sqrt{\left(\frac{\de z_{ij}}{ H(\bar z)}\right)^2\!\! + \left(\frac{2 \pi}{\ell}r(\bar z)\right)^2} & \mbox{if}  ~ 0 \leq \de z_{ij} \leq H(\bar{z}) \lambda_{\min}  \,.
 \end{array}  \right.
\eea
The highest multipole fulfilling these inequalities is
\be \label{lmaxdef}
\ell_{\max}^{ij} = \left\{\begin{array}{ll}
2 \pi r(\bar z)/\la_{\min}  & \mbox{ if } \de z_{ij}\le 0 \,, \\
\frac{2 \pi r(\bar z)}{\sqrt{\la_{\min}^2 - \left(\frac{\de z_{ij}}{ H(\bar z)}\right)^2}} & \mbox{ if }
0< \frac{\de z_{ij}}{ H(\bar z)}<\la_{\min}\,, \\
\infty & \text{ otherwise.}
\end{array} \right.
\ee
In the last case, the cut-off is given by the experimental angular resolution, $\ell_{\max}^{ij} = 2\pi/\theta_\mathrm{exp}$. Only multipoles $C_{\ell}^{ij}$ which satisfy the condition $\ell \leq \ell_{\max}^{ij}$ are taken into account in our analysis.
In the covariance matrix, for the spectra of redshift bins $(ij)$ and $(pq)$, we cut off the sum over multipoles in the Fisher matrix, Eq.~(\ref{eq:fisher}) at
\be \label{conservative}
\ell_{\max}={\rm min}(\ell_{\max}^{ij},\ell_{\max}^{pq})\,.
\ee
The nonlinearity scale is fixed by the smallest redshift difference appearing in the two pairs $(ij)$,  $(pq)$.
This condition ensures that nonlinear scales do not contribute to the derivatives in the Fisher matrix, Eq.~(\ref{eq:fisher}). This is an important limitation because nonlinear scales contain a large amount of information. Clearly, it will be necessary to overcome this limitation at least partially to profit maximally from future surveys. Notice that with $\ell_{\max}$ as in Eq.~(\ref{conservative}) the  the Fisher matrix can still involve non-linear contributions, but these are confined to the inverse of the
covariance matrix (\ref{e:cov}) where also the spectra $(iq)$, $(jp)$, $(ip)$ and $(jq)$ appear.
Of course, we need to use values for the cosmological parameters to determine  $r(z)$ and $H(z)$ in (\ref{e:lin}), but since this condition is approximate, this does not significantly compromise our results.

\section{Results}\label{s:res}

We now present Fisher matrix forecasts for several different types of surveys. In the error budget we only take into account sample variance, shot noise and photometric redshift (photo-$z$) uncertainties. In this sense, our results are not very realistic: the true analysis, containing also instrumental noise, is certainly more complicated. Nevertheless, we believe that this exercise is useful for the comparison of different methods and different surveys.

In what follows we refer to the $P(k)$ analysis as the 3D case, and to the $C_{\ell}$ analysis as the 2D case.
We assume a fiducial cosmology described by the minimal (flat) $\Lambda$CDM model, neglecting neutrino masses.
We study the dependence on the following set of cosmological parameters: ($\omega_{b}\equiv\Omega_bh^2$, $\omega_{CDM}\equiv\Omega_{CDM}h^2$, $H_0$, $A_s$, $n_s$), which denote the baryon and CDM density parameters, the Hubble parameter, the amplitude of scalar perturbation and the scalar spectral index, respectively.  We set the curvature $K=0$.

\begin{figure*}[tbp]
\centering
\includegraphics[width=0.45\textwidth]{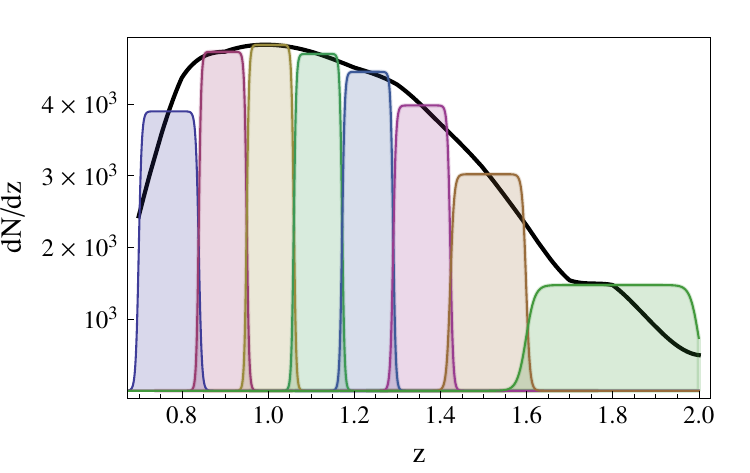}
\includegraphics[width=0.45\textwidth]{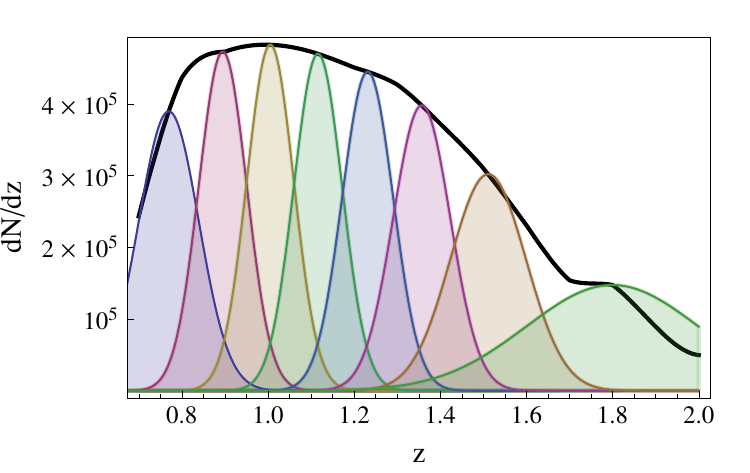}
\caption{Optimal binning strategy for an Euclid-like survey. The black line is the galaxy redshift distribution~\cite{EuclidRB}. We show the case in which the $z$-interval is divided into 8 bins. The width of every bin is chosen in order to have the same number of galaxies per bin. In the left panel we consider a spectroscopic survey, with tophat bins rounded at the edges to avoid numerical instabilities. (This rounding is not really necessary, but it allows us to use the same integration routine for both, tophat and Gaussian windows. Alternatively we can introduce hard limits in the case of tophat windows. We have compared both approaches and found that the difference is negligible.) On the right we show the case of a photometric survey, for which bins are chosen to be Gaussian because of photo-$z$ uncertainties.}
\label{f:redshift_binning_Euclid}
\end{figure*}

\subsection{A Euclid-like catalog}
In this section we perform a Fisher matrix analysis for an Euclid-like catalog. We compare the galaxy surveys with spectral and photometric redshifts.  We show the dependence  of the FoM on  redshift binning and on the nonlinearity scale. We concentrate on a few crucial observables since this is an illustration of our method and not a comprehensive forecast of a specific experiment. We defer to future work the inclusion of systematic errors and instrumental noise for a given experiment.

\subsubsection*{Methodology}

We determine the FoM for the joint estimation of the Hubble parameter $H_0$ and of the CDM density parameter $\omega_{\text{CDM}}$, marginalized over the other cosmological parameters ($\omega_{b}$, $A_s$, $n_s$). We do the same for the joint estimation of the baryon density parameter $\omega_b$ and $\omega_{\text{CDM}}$, marginalized over ($H_0$, $A_s$, $n_s$). We adopt two different non-linearity scales, $\lambda_{\text{min}} = 34 \ \text{Mpc}/h$ and $\lambda_{\text{min}} = 68 \ \text{Mpc}/h$.
We assume an angular resolution of about $3$ arc minutes such that in addition to the condition given in Eqs.~(\ref{lmaxdef}, \ref{conservative}) we impose $\ell_{\max}\leq3000$.
Note that we do not combine with known datasets like Planck to minimize the uncertainties on parameters. The goal here is not an optimal prediction of the improvement on cosmological parameter estimation by Euclid (for this we would need a detailed treatment of instrument errors). Here we want to compare two methods,
 the 3D and 2D analysis. In addition, we also study the importance of off--diagonal correlators and of the non-linearity scale.

\subsubsection*{Binning strategy}
The overall galaxy redshift distribution of the survey in the considered range $0.7<z<2$ is the black line in Fig.~\ref{f:redshift_binning_Euclid}, see~\cite{EuclidRB}\footnote{See also \url{http://www.euclid-ec.org/}.}.
Our binning strategy is to adjust the width of each bin such that there is the same number of galaxies per redshift bin. We verified that this choice gives higher FoM's than the choice of constant bin width, because it minimizes  shot noise.

We consider a spectroscopic and a photometric survey. In the first case bins are tophat in redshift. In practice, to avoid numerical instabilities, \classgal{} uses rounded edges, see~\cite{Didio1}. For the photometric case, we assume to dispose of 100 times more galaxies, but to take into account the uncertainties of the redshifts, we model the bins as Gaussian. The width of the bins must be consistent with the photo-$z$ errors $0.05(1+z)$, for  a splitting of the redshift interval $0.7<z<2$ this corresponds to about 6 bins. Instead, spectroscopic precision $0.001(1+z)$ is not an issue in our analysis.

In both cases (spectroscopic and photometric), we compare the results of a 2D and 3D analysis.
For the 3D spectroscopic case, the FoM is nearly independent of the number of bins, within a few percent. Hence in the following plots we adopt a constant value computed with $N_\text{bin}=1$.  The reason for this is that we assume a known redshift dependence of the growth factor, $f(z) = \Om_M(z)^\ga$ with $\ga =0.6$. If we would want to determine $\ga$ or to reconstruct the bias, $b(z)$, increasing the number of bins would be significant also for the 3D analysis. However, since this aspect  is very similar for both, the 2D and the 3D analysis, we do not study it in this work.

In the 3D photometric case, the FoM is no longer constant, but rather decreases slowly with $N_\text{bin}$. In fact, photometric uncertainties cause redshift bins to be correlated, hence the total Fisher matrix is not well approximated by the sum of the Fisher matrices of each bin, as assumed in \cite{Asorey+12} and in the present work. Since we wish to refer to the most favorable bin configuration in the 3D analysis, we will always compare 2D photometric results with the 3D photometric FoM obtained for $N_\text{bin}=1$.

\begin{figure*}[tbp]
\centering
  \hspace*{0.5cm} $\la_{\min}= 34h^{-1}$Mpc \hspace{3.9cm} $\la_{\min}= 68h^{-1}$Mpc\\
\includegraphics[width=0.45\textwidth]{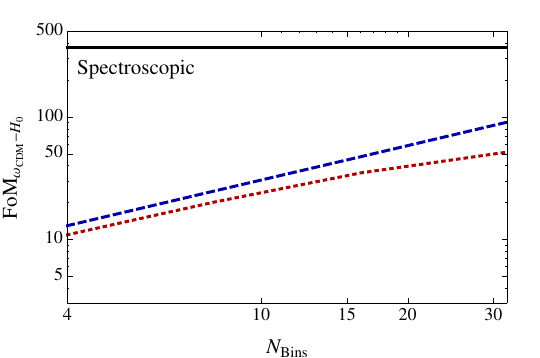}
\includegraphics[width=0.45\textwidth]{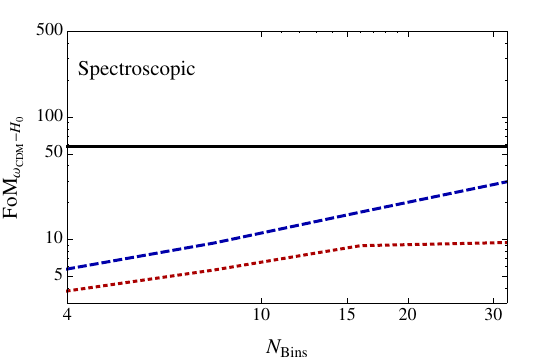} \\
\includegraphics[width=0.45\textwidth]{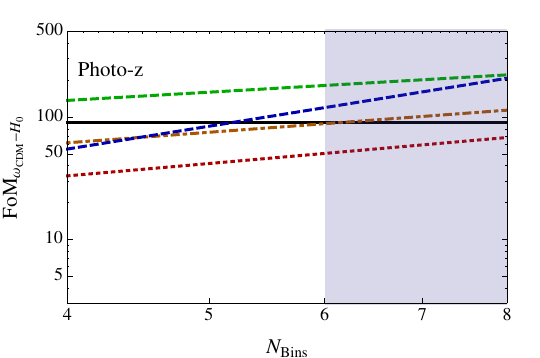}
\includegraphics[width=0.45\textwidth]{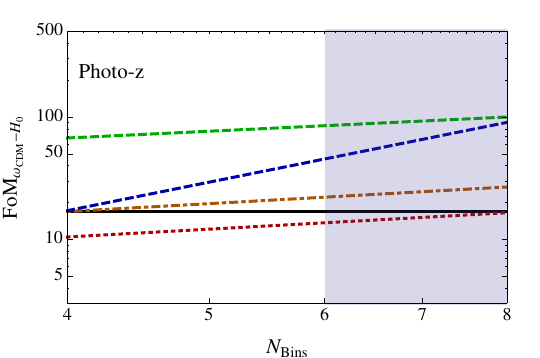}
\caption{FoM for an Euclid-like survey for the non-linearity scales $34\ \text{Mpc}/h$, left panels, and $68\ \text{Mpc}/h$, right panels.
Top panels refer to spectroscopic redshifts, bottom panels to photometric ones.
For the 2D FoM we consider a cut at $\ell_\text{max}$ values defined in Eq.~(\ref{conservative}) (dashed blue), and also the case in which cross-correlations of bins are neglected (dotted red). For the photometric survey, we  plot the result derived from two different definitions of the bin separation (and hence of $\ell_\text{max}$), defining ($z_i^{\rm sup}$, $z_j^{\rm inf}$) at either $2$- (green) or $3$-$\si$ (blue), see  Footnote~\ref{two_or_three}. For this case we also plot the FoM for the $\ell_\text{max}$ determination of Ref.~\cite{Asorey+12} (dot-dashed orange).
The solid black line shows the value of the 3D FoM for $N_\text{bin}=1$ and $\sigma_z=0.05(1+\bar z)$ in the photometric case. The shadowed region exceeds this $z$ resolution.
}
\label{f:FoM}
\end{figure*}

\subsubsection*{2D and 3D performances}

Not surprisingly, considering a smaller nonlinearity scale $\lambda_\text{min}$ yields a larger FoM (compare the left and right panels of Figure \ref{f:FoM}). Naively, we expect to gain a factor 8 in the FoM when going from $\la_{\min}=68h^{-1}$Mpc to $\la_{\min}=34h^{-1}$Mpc, since we have $2^3$ times more Fourier modes at our disposition.
Fig.~\ref{f:FoM} quite accurately confirms this expectation for the 3D case (black lines).
Interestingly, the 2D FoM degrades only by about a factor of 2 when the non-linearity
scale is increased by 2. Since the redshift resolution is not affected by $\la_{\min}$,  naively we would expect a degradation by a factor of 4 from the transverse directions, but it seems that the redshift information is as important as both transverse direction so that
we only loose a factor of 2.

In Figure \ref{f:FoM}, we also show the FoM adopting different definitions of $\ell_{\max}$ and including only auto-correlations of redshift bins.  Apart from our definition of $\ell_{\max}$ given in Eqs.~(\ref{lmaxdef}, \ref{conservative}), we
 also consider the definition $\ell_{\max}=r(\bar z) 2\pi/\la_{\min}$ used in \cite{Asorey+12}, where $r(\bar z)$ is the comoving distance to the mean redshift $\bar z=1.35$ of the survey. While in the spectroscopic case this yields results which are within 30\% of those obtained with our method, in the photometric case the FoM is significantly improved when using our definition, compare the (dot-dashed) orange and the (long dashed) green lines in  the lower panels of Figure \ref{f:FoM}. The definition of Ref.~\cite{Asorey+12} is sufficient for a narrow band survey, but not for a Euclid-like survey.
This shows how much information can be gained by using  our optimal definition of $\ell_{\max}$ given in Eq.~(\ref{conservative})  but also by taking into account cross-correlations (dashed blue lines) as compared to considering only the auto-correlation of redshift bins (dotted red lines). For photometric redshifts the result strongly depends on the value chosen for $\de z_{ij}$ which determines $\ell_{\max}^{ij}$. The larger  is
$\de z_{ij}$, the more radial information is lost. Compare the green (dashed, top) and blue (dashed, lower) lines for the photo-z FoM's, where we compare the choice of 2 and 3 standard deviation for $\de z_{ij}$. This is the problem of Gaussian binning. It does spread  the radial information considerably. Nevertheless, the reduction of shot noise (due to the fact that the photometric survey is assumed to contain a hundred times more galaxies) compensates for this, and leads to a better FoM from photometric surveys in the 2D analysis. However, in the spectroscopic case we could still increase the number of bins.

Interestingly, with the more conservative value of the non-linearity scale $\lambda_\mathrm{min}= 68\ \text{Mpc}/h$ (right hand panels of Fig.~\ref{f:FoM}), the difference between the full analysis (blue, dashed) and the one involving only auto-correlations (red, dotted) becomes very significant also for spectroscopic surveys. When considering only auto-correlations, the spectroscopic FoM reaches saturation at 16 bins.
One reason for this is that shot noise affects auto-correlations more strongly than cross-correlations. But even neglecting shot noise, we still observe this saturation.  We just cannot gain more information from the auto-correlations alone by increasing the number of bins, i.e. by a finer sampling of transverse correlations.

Comparing the spectroscopic with the photometric survey and sticking to the 2D analysis,  we find that the photometric specifications yield a larger FoM. This is not surprising, since the latter case contains 100 times more galaxies than the spectroscopic survey, for which shot noise is correspondingly 100 times larger.
Also, the FoM for the 2D analysis mainly comes from cross correlations. This is why different choices of $\de z_{ij}$ affect the final result so much.

The photometric FoM from the 3D analysis, however, is lower than the one from the spectroscopic survey. In the 3D case, the redshift uncertainties translate into uncertainties in the wavenumber $k$ which are more relevant than the reduction of shot noise.

When the number of bins is large enough, our 2D analysis yields a better FoM than the standard $P(k)$ analysis. For the photometric survey this is achieved already at $N_{\rm bin}=4-6$ while for the spectroscopic survey probably about 120 bins would be needed.

At some maximal number of bins the number of galaxies per bin becomes too small and shot noise starts to dominate. At this point  nothing more can be gained from increasing the number of bins. However, since for slimmer redshift bins not only the shot noise but also the signal increases, see~\cite{Didio1} for details, the optimal number of bins is larger than a naive estimate. In Ref.~\cite{Didio1} it is shown that the shot noise, which behaves like $\ell(\ell+1)/2\pi$ (in a plot of $\ell(\ell+1)C_\ell /2\pi$),
becomes of the order of the signal at somewhat smaller $\ell$ for slimmer redshift bins.
Again, due to the increase of the signal this dependence is rather weak down to a redshift width of $\De z = 0.0065$.
For redshift slices with $\De z < 0.005$, the signal does not increase anymore while the shot noise still does. For a DES-like survey, this maximal number of bins is about $N_{\max}^{\rm (DES)}\sim 50$, while for a Euclid-like survey it is of the order of
$N_{\max}^{\rm (Euclid)}\sim 200$.

When using redshift bins which are significantly thicker than the redshift resolution of the survey, the 3D analysis, in principle, has an advantage since it makes use of the full redshift resolution in determining distances of galaxies, while in the 2D analysis we do not distinguish redshifts of galaxies in the same bin. A  redshift bin width of the order of the nonlinearity scale beyond which the power spectrum is not reliable, given by
$r(\bar z, \De z) \sim 2\De z/ H(\bar z) \simeq\la_{\min}$ is the minimum needed to recover the 3D FoM for spectroscopic surveys.

 However, for spectroscopic surveys we can in principle allow for very slim bins with a thickness significantly smaller than the nonlinearity scale, and the maximal number of useful bins is decided by the shot noise.  Comparing the $C_\ell^{ii}$ with $C_\ell^{jj}$, e.g., of neighboring bins still contains some information, e.g., on the growth factor, even if $\ell_{\max}^{ij} \sim \ell_{\max}^{ii}$ for small $|i-j|$. In simpler terms, the fact that our analysis effectively splits transversal information coming from a given redshift and radial information from $C_\ell^{ij}$ with large $\ell$, makes it in principle advantageous over the $P(k)$ analysis. It becomes clear from Fig.~\ref{f:FoM} (see also Fig.~\ref{f:2Dvs3D}  below) that we need to use sufficiently many bins and our definition of $\ell_{\max}^{ij} $ to fully profit from this advantage.

The numerical effort of a Markov Chain Monte Carlo analysis of real data scales roughly like $N_{\rm bin}^2$. Running a full chain of, say $10^5$ points in parameter space requires the calculation of about $10^5N_{\rm bin}^2/2$  spectra with \class{}. (For comparison, a CMB chain requires `only' $4\times10^5$ spectra).

Note also, that the advantage of the 2D method is relevant only when we want to estimate cosmological parameters. If the cosmology is assumed to be known and we want, e.g., to reconstruct the bias of galaxies w.r.t.~matter density fluctuations, both methods are equivalent. But then, the redshift dependence also of $P(k)$ has to be studied and both  power spectra, $P(k,\mu,z)$ and $C_\ell(z,z')$ are functions of three variables.

\begin{figure*}[tbp]
\centering
\includegraphics[width=0.45\textwidth]{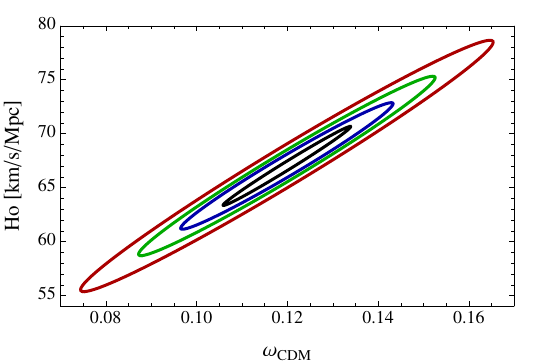}
\includegraphics[width=0.45\textwidth]{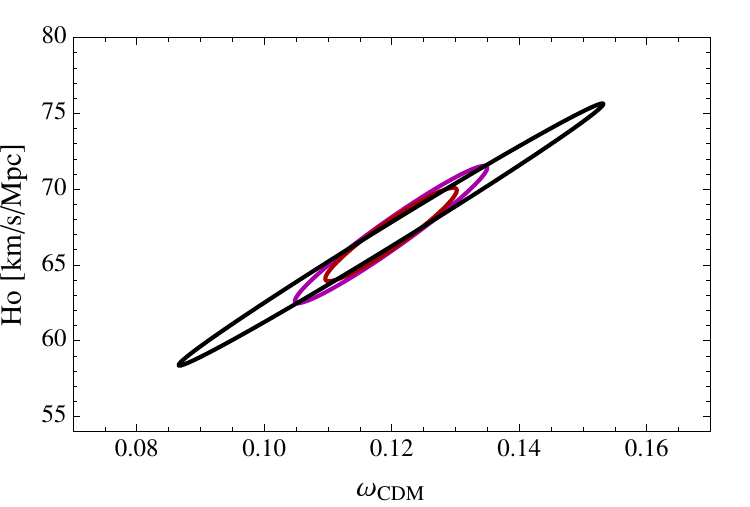}
\caption{Confidence ellipses in the $H_0\!-\!\Omega_\text{CDM}h$ plane (marginalized over the other cosmological parameters) for an Euclid-like survey computed with a non-linearity scale of $34\ \text{Mpc}/h$. Different colors indicate different redshift bin numbers: 4 bins (magenta), 8 bins (red), 16 bins (green), 32 bins (blue). Black contours refer to the 3D analysis for $N_\text{bin}=1$. We consider a spectroscopic survey (left panel) and a photometric one (right panel, for $2\si$ bin width).
}
\label{f:Ellipse_Euclid}
\end{figure*}

In Figure \ref{f:Ellipse_Euclid} we plot, as one of several possible examples, the confidence ellipses in the $H_0\!-\!\omega_\text{CDM} $ plane (marginalized over the other cosmological parameters) for an Euclid-like survey computed with a nonlinearity scale of $34\ \text{Mpc}/h$ for different redshift bins. We do not assume any prior knowledge e.g.~from Planck. Therefore again, this is not a competitive parameter estimation but only a comparison of methods. We also show the dependence of the result on the number of bins.
The strong degeneracy between $H_0$ and $\om_{\rm CDM}$ comes from the fact that $\om_{\rm CDM}$ is mainly determined by the break in the power spectrum at the equality scale $k_{\rm eq}$ which of course also depends on $H_0$. As is well known, the break in the power spectrum actually determines $\Om_{\rm CDM}h \propto \om_{\rm CDM}/H_0$. This determines the slope of the ellipse in Figure \ref{f:Ellipse_Euclid} which is extremely well constrained. Note also that the 2D and 3D ellipses have a slightly different slope, hence different principal axis. This implies that they do not constrain exactly the same combinations of $H_0$ and $\om_{\rm CDM}$.

\begin{figure*}[tbp]
\centering
\includegraphics[width=0.45\textwidth]{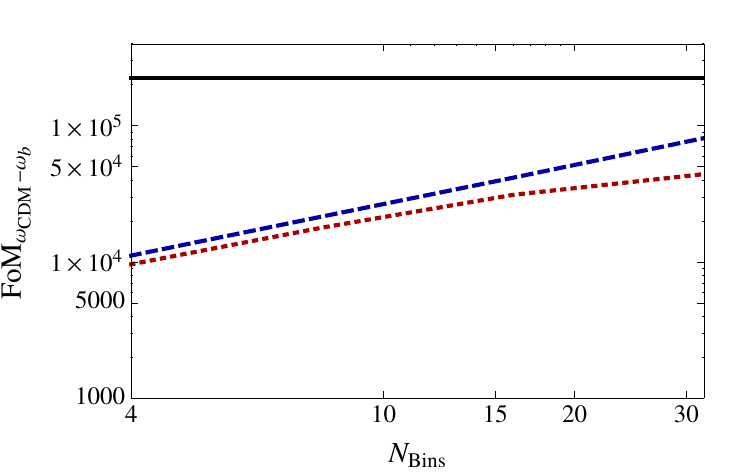}
\includegraphics[width=0.45\textwidth]{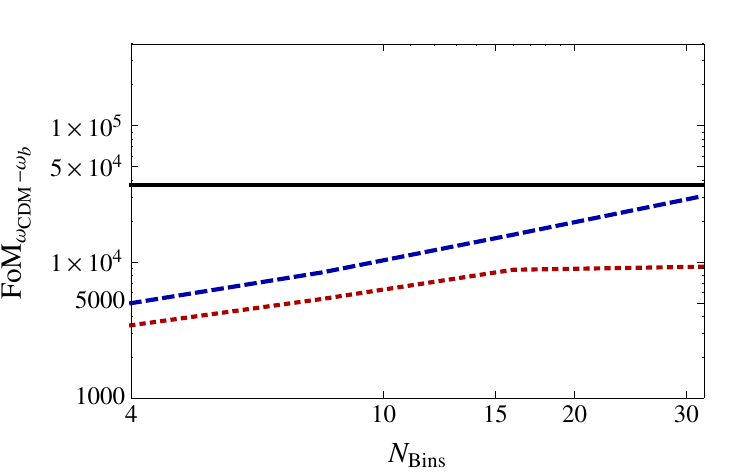}
\caption{FoM for $\om_\text{CDM} $ - $\omega_b$ (marginalized over other cosmological parameters) for different nonlinearity scales ($34\ \text{Mpc}/h$, left panels, and $68\ \text{Mpc}/h$, right panels) for a spectroscopic Euclid-like survey.
For the 2D FoM we consider $\ell_\text{max}$ as defined in Eq.~(\ref{conservative}) (dashed blue), and also the case in which cross-correlations of $z$-bins are neglected (dotted red).
The solid black line shows the value of the 3D FoM for $N_\text{bin}=1$.}
\label{f:FoM_Euclid_omB}
\end{figure*}

In Fig.~\ref{f:FoM_Euclid_omB} we show the FoM  for $(\om_b,\om_{\rm CDM})$ for the spectroscopic Euclid survey. We marginalize over the other parameters, $(H_0,n_s,A_s)$.
Also in this case, the important contribution from cross--correlations, especially for the larger non-linearity scale is evident.

\subsection{A DES-like catalog}
We perform the same analysis as in the previous section also for a DES-like survey\footnote{\url{www.darkenergysurvey.org}}.  A similar analysis has been performed in  Refs.~\cite{Asorey+12,Asorey:2013una}, and the goal of this section is  to compare our results with these references.  The novelties of the present analysis is that we marginalize over cosmological parameters which have been fixed in Refs.~\cite{Asorey+12,Asorey:2013una}  and we use a more sophisticated definition of the nonlinearity scale.
Following~\cite{Asorey+12} we consider a galaxy density distribution $dN/dz$ given by
\be
\frac{dN}{dz} \propto \bra{\frac{z}{0.55}}^2 \exp\sbra{-\bra{\frac{z}{0.55}}^2} ~,
\ee
in the redshift range $0.45<z<0.65$, which is divided in $z$-bins of the same size. The shot noise is determined by the galaxy number density, assumed to be $n=3.14 \times 10^{-3} h^3 \text{Mpc}^{-3}$, constant in $z$. In this case we only consider a spectroscopic redshift resolution.

\begin{figure*}[tbp]
\centering
\includegraphics[width=0.45\textwidth]{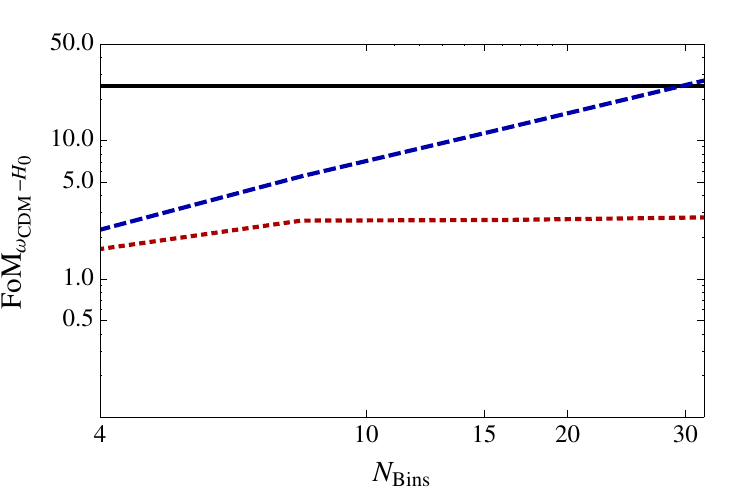}
\includegraphics[width=0.45\textwidth]{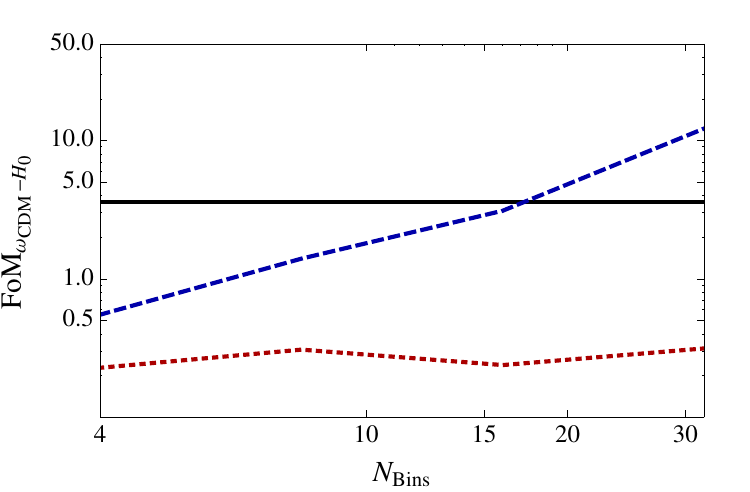} \\
\caption{FoM for the nonlinearity scales $34\ \text{Mpc}/h$, left panels, and $68\ \text{Mpc}/h$, right panels for a spectroscopic, DES-like survey.
For the 2D FoM we consider $\ell_\text{max}$ as defined in Eq.~(\ref{conservative}) (dashed blue), and also the case in which cross-correlations between bins are neglected (dotted red).
The solid black line shows the value of the 3D FoM for $N_\text{bin}=1$.
}
\label{f:FoM_DES}
\end{figure*}

In Fig.~\ref{f:FoM_DES} we show the FoM for the determination of $(H_0,\omega_\text{CDM})$ from a DES-like survey, marginalizing over the remaining cosmological parameters. In this case, the FoM coming from autocorrelations alone saturates already at $N_\text{bin}=8$. The FoM including cross-correlations continues to grow and overtakes
the one from the 3D analysis at $N_\text{bin}=30$.

\begin{figure*}[tbp]
\centering
\includegraphics[width=0.45\textwidth]{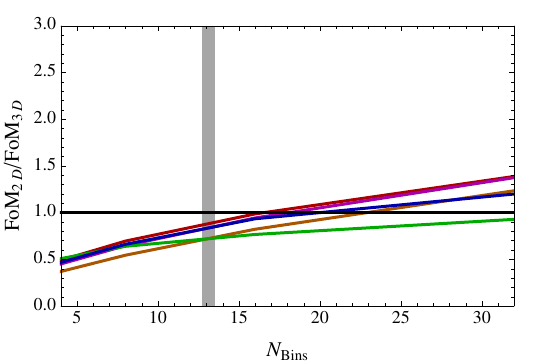}
\includegraphics[width=0.45\textwidth]{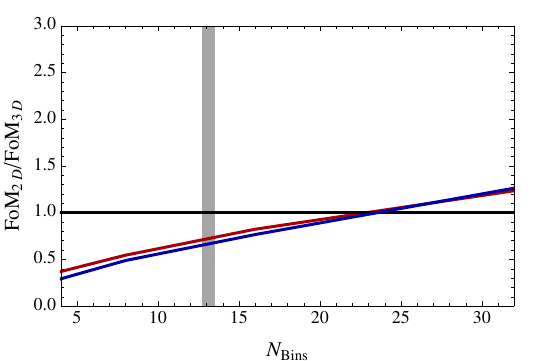}
\includegraphics[width=0.45\textwidth]{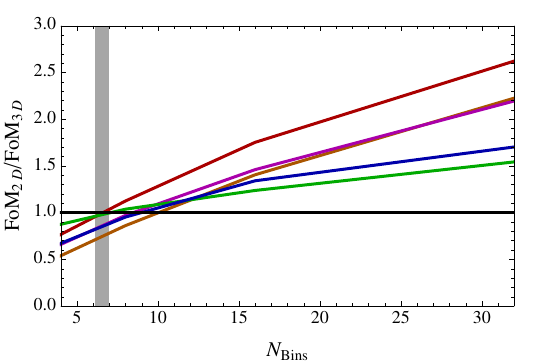}
\includegraphics[width=0.45\textwidth]{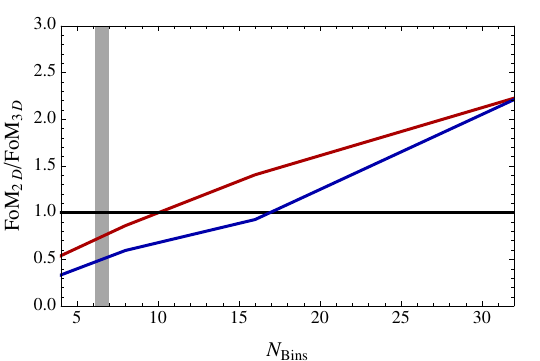}
\caption{In the left panels we show the ratio between FoM 2D over 3D for each single parameter keeping the others fixed for a DES-like survey. Different colors indicate different parameters: $\om_b$ (red), $\om_{\rm CDM}$ (orange), $n_s$ (magenta), $H_0$ (green) and $A_s$ (blue). The vertical thick gray line shows when the bin size becomes of the same order of
the non-linearity scale, where we expect that 2D and 3D analysis should roughly give the same result. This seems only approximately true.
On the right panels we show the ratio between FoM's from  2D over 3D for $\om_{CDM}$. In red keeping the other cosmological parameters fixed, while in blue
marginalizing over all other parameters.
In the top panels we consider the nonlinearity scale $\la_{\min} = 34h^{-1}$Mpc while in the bottom  panels we have $\la_{\min} =68h^{-1}$Mpc.}
\label{f:2Dvs3D}
\end{figure*}

In Fig.~\ref{f:2Dvs3D} we compare the FoM of each parameter in the 2D and 3D case for different values of the non-linearity scale. The mean number of bins at which the nonlinearity scale is reached, $ \De z \simeq\la_{\min}H(\bar z)/2$, is indicated as vertical grey bar. The number of bins where the 2D analysis becomes better than the 3D one, especially for the marginalized FoM, is typically nearly a factor of 2 larger than this naive estimate. This may come from the fact that we  include correlations with small (or vanishing) angular separation only if $|z^{\rm inf}_j-z^{\rm sup}_i|$ is large enough while for most galaxies the mean bin distance $|\bar z_j-\bar z_i|$ would be relevant.
In this sense our choice is conservative.

Interestingly, comparing Euclid and DES FoM's we find that while the $P(k)$ FoM for a Euclid-like survey is more than a factor of  10 times better than the one of a DES-like survey, this is no longer true when we compare the FoM's from our angular analysis at fixed number of redshift bins. However, to find the true (optimal but still realistic) FoM we have to increase the number of redshift bins until the FoM converges to its maximum. In practice at some point the instrumental noise which has been neglected in our treatment will prevent further growth of the FoM so that we refrained from going beyond $N_{\rm bin}=32$ which would also be numerically too costly with our present implementation, but which will be interesting for a future analysis.

\begin{figure*}[tbp]
\centering
\includegraphics[width=0.45\textwidth]{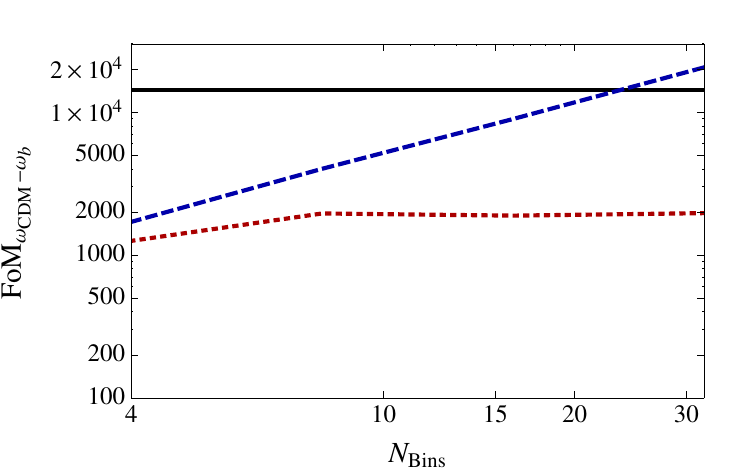}
\includegraphics[width=0.45\textwidth]{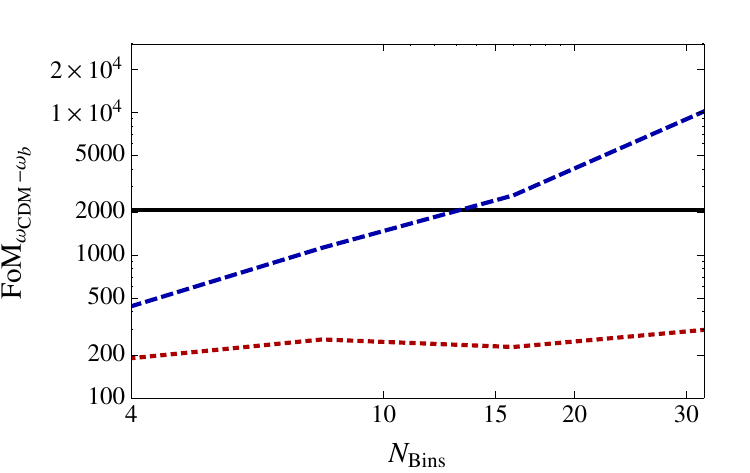}
\caption{FoM for $\omega_\text{CDM}$ - $\omega_b$ (marginalized over other cosmological parameters) for different nonlinearity scales ($34\ \text{Mpc}/h$, left panel, and $68\ \text{Mpc}/h$, right panel) for a spectroscopic DES-like survey.
For the 2D FoM we consider $\ell_\text{max}$ as defined in Eq.~(\ref{conservative}) (dashed blue), and also the case in which cross-correlations of $z$-bins are neglected (dotted red).
The solid black line shows the value of the 3D FoM for $N_\text{bin}=1$.
}
\label{f:FoM_DES_omB}
\end{figure*}

For completeness, we also show the FoM  for $(\om_b,\om_{\rm CDM})$ for the DES like surveys marginalizing over $(H_0,n_s,A_s)$ in Fig.~\ref{f:FoM_DES_omB}.  Again, the FoM from auto-correlations only (red, dotted) saturated at about 8 redshift bins.

\subsection{Measuring the lensing potential}\label{s:lens}
It is well known that the  measurement of the  growth rate  requires an analysis
of  galaxy surveys which are sensitive to redshift-space distortions~\cite{Ross:2007,Reid:2012sw,Contreras:2013,delaTorre:2013}.
Isolating other effects can lead to an analysis which is more sensitive to other parameters.
In this section we study especially how one can measure the lensing potential with galaxy surveys. The lensing potential is especially sensitive to theories of modified gravity which often have a different lensing potential than General Relativity, see, e.g.~\cite{Asaba:2013xql,Saltas:2010tt,Durrer:2007re}.  The lensing potential out to some redshift $z$ is defined by~\cite{Durrer:2008aa}
\be
\varPsi_\ka(\bn,z) = \int_0^{r_s(z)}dr\frac{r_s-r}{r_sr}\left(\Psi(r\bn,t)+\Phi(r\bn,t)\right) \,.
\ee
Denoting its power spectrum by $C_\ell^{\varPsi}(z,z')$, we can relate it to the lensing contribution $C_\ell^{\rm lens}(z,z')$ \cite{Didio1} to the angular matter power spectrum $C_\ell(z,z')$ by
\be
C_\ell^{\rm lens}(z,z') = \ell^2(\ell+1)^2C_\ell^{\varPsi}(z,z') \,.
\ee
We shall see, that this lensing power spectrum can be measured from redshift integrated angular power spectra of galaxy surveys.

To study this possibility, we first introduce the signal-to-noise for the different terms which contribute to the galaxy power spectrum as defined in Ref.~\cite{Didio1}. For completeness we list these terms in Appendix~\ref{appA}. The  signal-to-noise for a given term is given by
\be
\label{e:SN}
\left(\frac{S}{N}\right)_\ell = \frac{\left| C_{\ell}-\tilde{C}_{\ell}\right|}{\sigma_{\ell}} \;.
\ee
where $\tilde{C}_{\ell}$ is calculated neglecting the term under consideration (e.g., lensing), and the r.m.s. variance is given by
\be
\label{e:CosmVar}
\sigma_{\ell} = \sqrt\frac{2}{(2\ell+1)f_{\rm sky}} \bra{ C_{\ell} + \frac{1}{n}}  \;.
\ee
It is also useful to introduce a cumulative signal-to-noise that decides whether  a term is observable within a given multipole band. We define the cumulative  signal-to-noise by
\be
\label{e:SNcum}
\bra{ \frac{S}{N} }^2 = \sum_{\ell=2}^{\ell_{\max}} \bra{\frac{C_{\ell}-\tilde{C}_{\ell}}{\sigma_{\ell}}}^2\,.
\ee
Note that $C_{\ell}-\tilde{C}_{\ell}$ contains not only the auto-correlation of a given term, but also its cross-correlations with other terms so that it can be negative. Especially, for small $\ell$'s the lensing term is dominated by its anti-correlation with the density term and is therefore negative.

Eq.~(\ref{e:SN}) estimates the contribution of each term to the total signal.
If its signal-to-noise is larger than 1, in principle it is possible measure this term and therefore to constrain cosmological variables determined by it.
To evaluate the signal-to-noise of the total $C_{\ell}$'s, which is the truly observed quantity, we set $\tilde{C}_{\ell}=0$.

\begin{figure*}[tbp]
\centering
\includegraphics[width=0.48\textwidth]{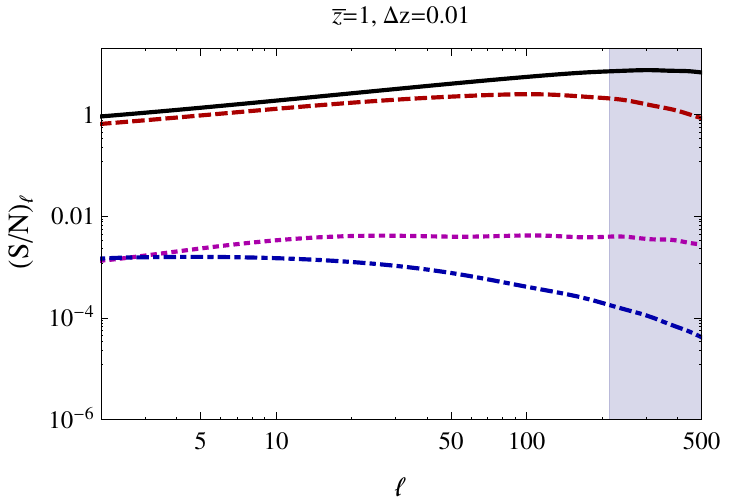}\\
\includegraphics[width=0.48\textwidth]{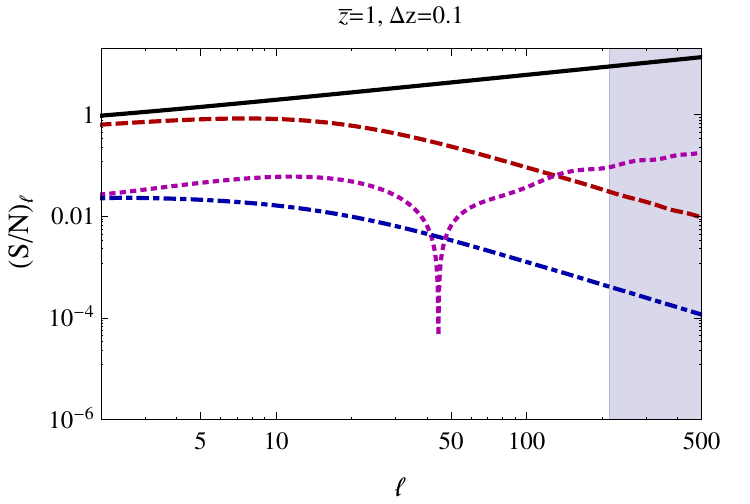}
\includegraphics[width=0.48\textwidth]{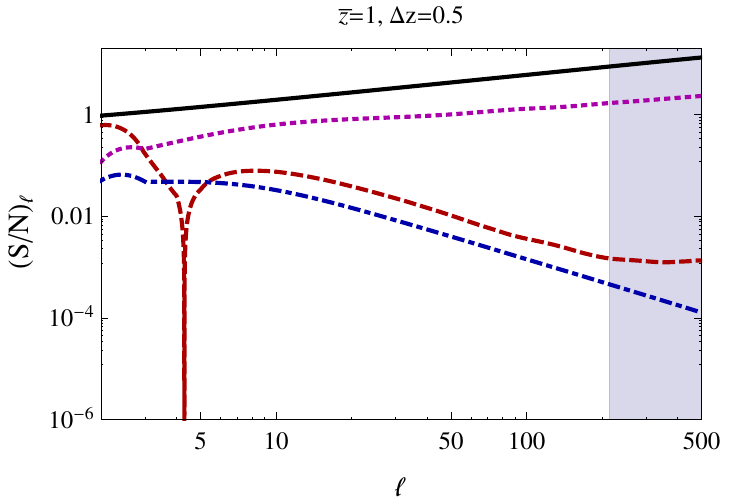}\\
\includegraphics[width=0.48\textwidth]{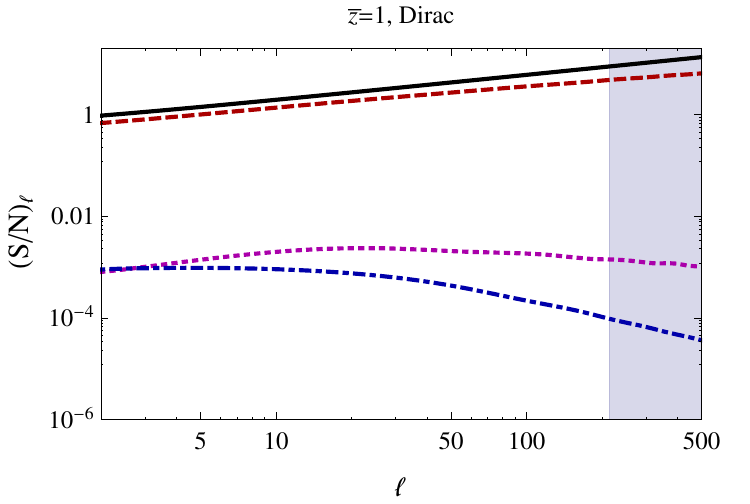}
\includegraphics[width=0.48\textwidth]{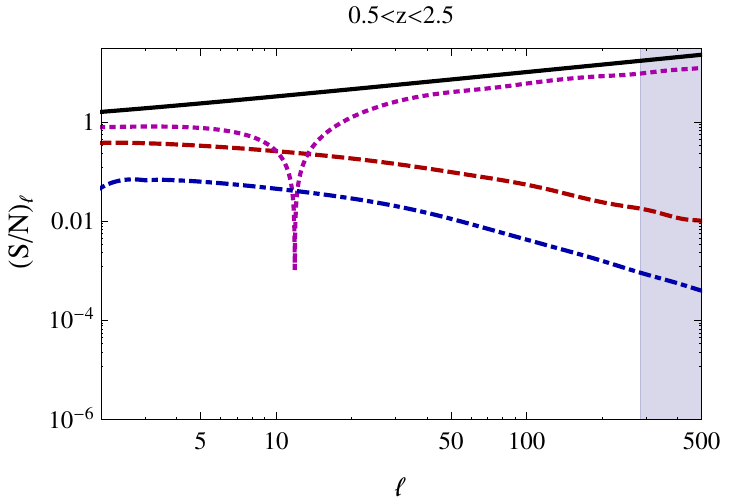}
\caption{Signal-to-noise for different terms: total (solid, black), redshift-space distortions (dashed, red), lensing (dotted, magenta), potential terms (dot-dashed, blue).
The plot in the top line is computed with a tophat window function with half-width $\De z=0.01$.
The  cases in the second line correspond to Gaussian window functions of half-width $\De z=0.1,0.5$ around $\bar{z}=1$.
These three plots are compatible with Euclid specifications.
In the third line, we show two extreme situations.
The Dirac z-window function corresponds to an infinitesimal $z$-bin (left) and a very
wide redshift range, $0.5<z<2.5$. This last case shows a large lensing signal. Notice the zero-crossing of the lensing term which is indicated by a downward spike in this log-plot.
The shadowed regions correspond to nonlinearity scale estimated at the mean redshift of the bin.}
\label{f:SN_terms}
\end{figure*}

In Figure \ref{f:SN_terms} we show the signal-to-noise for different width of the redshift window function.
We consider a tophat window for the narrowest case, $\De z=0.01$ and Gaussian window functions with standard deviations $\De z\gtrsim0.05(1+z)$, which corresponds to Euclid photometric errors \cite{EuclidRB} for the panels on the second line. The sky fraction, $f_{sky}$, and the galaxy distribution $dN/dz$ are compatible with Euclid specifications (see black lines in Fig. (\ref{f:redshift_binning_Euclid})). In particular, shot noise turns out to be negligible in this analysis. The shadowed regions in Figure \ref{f:SN_terms} show the nonlinearity scales, estimated as $\ell_{\max}=2\pi r(\bar{z})/\lambda_{\min}$, where $\bar z$ is the mean redshift of the bin and $\lambda_{\min}=68h^{-1}$ Mpc.

As expected \cite{Bonvin:2011bg}, redshift-space distortions and purely relativistic terms are mainly important at large scales, while lensing has a weaker scale dependence.
For small $\De z$, apart from the usual density term, redshift-space distortions are the main contribution. Their signal-to-noise is larger than one, which allows to constrain the  growth factor.
As $\De z$ increases, redshift-space distortions are washed out, and their signal decreases significantly.
On the other hand, lensing and potential terms increase.
This is due to the fact that these terms depend on integrals over $z$ that
coherently grow as the width of the $z$-window function increases.
While potential terms always remain sub-dominant, the lensing signal-to-noise becomes larger than 1 for the  value of $ \De z=0.5$ already at $\ell\approx60$.

As a reference, we also show the signal-to-noise for an infinitesimal bin width (Dirac $z$-window function).
This corresponds to the largest possible $C_{\ell}$ amplitude.
As in the case $\De z=0.01$, redshift-space distortion is of the same order as the density term.
Note , however, that in reality  for very narrow bins, shot noise becomes important and decreases $(S/N)_{\ell}$, especially for large multipoles.

The case of a uniform galaxy distribution between $0.5<z<2.5$ is also shown.
We assume $f_{\rm sky}=1$ and neglect shot noise.
In this configuration the lensing term has a very large signal-to-noise.
This can be used to constrain the lensing potential by comparing the observable
(total) $C_{\ell}$'s to the theoretical models.
In practice one may adapt this study to catalogs of radio galaxies, which usually
cover wide $z$ ranges but with poor redshift determination which is not needed for this case, for previous studies see, e.g.,
\cite{Blake:2003dv,Negrello:2007kv,Subrahmanyan:2008au}.

\begin{figure*}[tbp]
\centering
\includegraphics[width=0.48\textwidth]{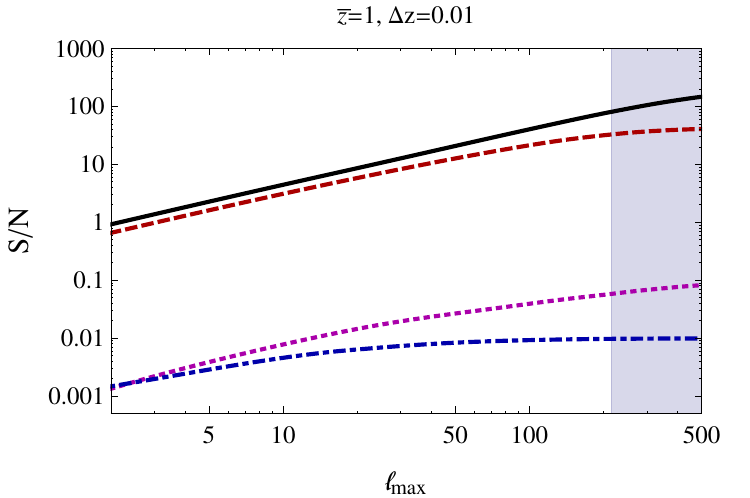}\\
\includegraphics[width=0.47\textwidth]{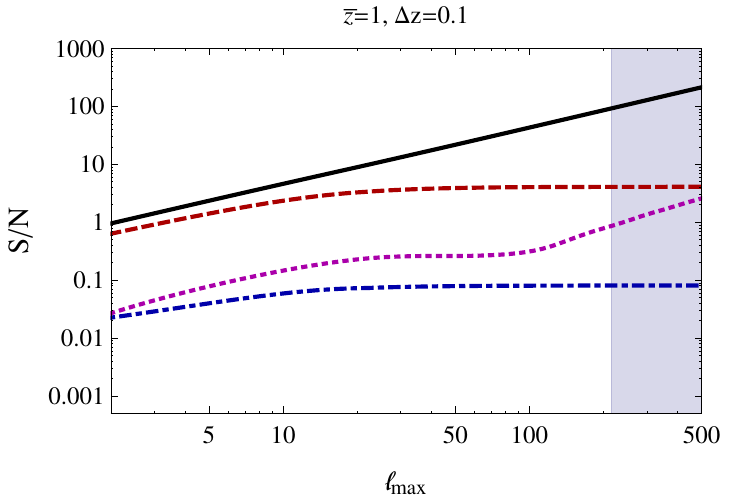}
\includegraphics[width=0.47\textwidth]{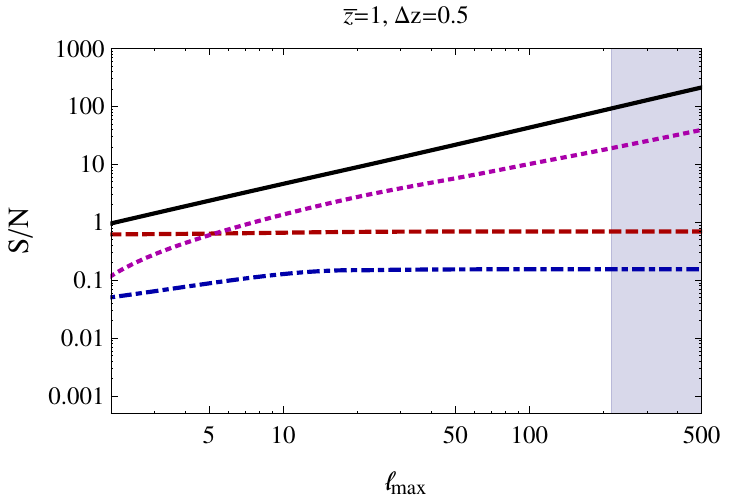}\\
\includegraphics[width=0.47\textwidth]{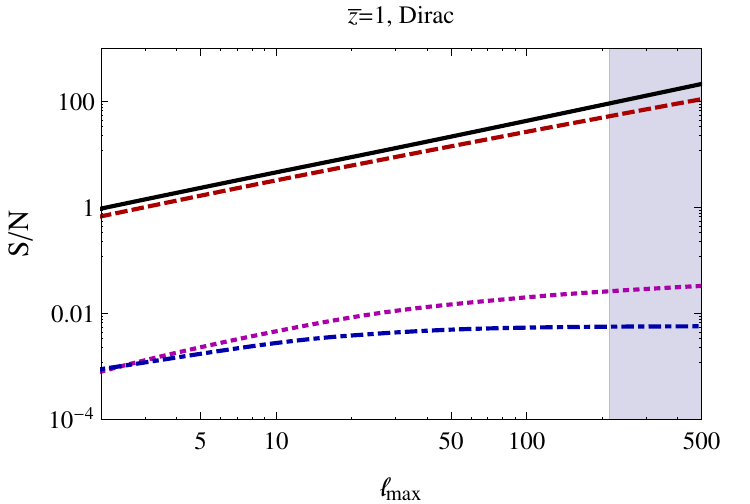}
\includegraphics[width=0.47\textwidth]{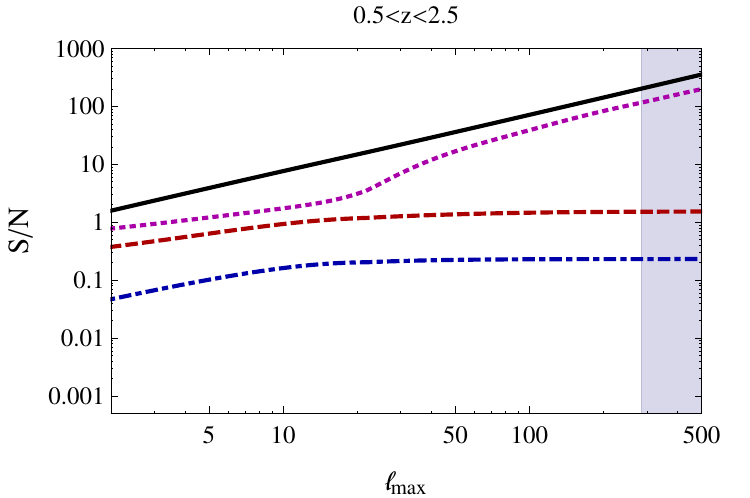}
\caption{
Cumulative signal-to-noise for different terms: total (solid, black), redshift-space distortions (dashed, red), lensing (dotted, magenta), potential terms (dot-dashed, blue).
The different plots correspond to the same configurations as in Figure \ref{f:SN_terms}.}
\label{f:SNcum_terms}
\end{figure*}

In Figure \ref{f:SNcum_terms} the cumulative signal-to-noise, Eq. (\ref{e:SNcum}), is shown as function of the maximum multipole considered in the sum.
Contrary to all the other terms, the cumulative signal-to-noise of the potential terms never exceeds 1. We therefore conclude that the considered experiment is not able to measure the potential terms. It is not clear, whether another feasible configuration would be sensitive to them. Notice also how the lensing term really 'kicks in' after the zero-crossing, when it is not longer dominated by its anti-correlation with the density term but by the contribution from the autocorrelation.

\subsection{The correlation function and the monopole}
So far we have not considered the observed monopole, $C_0^{ij}$, and dipole, $C_1^{ij}$, since the former is affected by the value of the gravitational potential and the density fluctuation at the observer position, while the latter depends on the observer velocity. These quantities are not of interest for cosmology and cannot be determined within linear perturbation theory.

However, there is an additional point which has to be taken into account when considering  the correlation function. Usually, the correlation function is determined from an observed sample of galaxies by subtracting from the number of pairs with given redshifts and angular separation the corresponding number for a synthetic, uncorrelated sample with the same observational characteristics (survey geometry, redshift distribution, etc.).
This is the basis of the widely used Landy--Szalay estimator for the determination of the correlation function from an observed catalogue of galaxies~\cite{Landy:1993}. For this estimator, by construction, the integral over angles vanishes, so that
\be
C_0^{({\rm LS})}(z_i,z_j) = 2\pi\int \xi_{\rm LS}(\theta,z_i,z_j)\sin\theta d\theta  =0 \,.
\ee
Here $ \xi_{\rm LS}(\theta,z_i,z_j)$ is already convolved with the redshift window function of the survey. If we want to compute the Landy--Szalay estimator for the correlation function we therefore have to subtract the monopole,
$$
 \xi_{\rm LS}(\theta,z_i,z_j) =  \xi(\theta,z_i,z_j) - \frac{1}{4\pi}C_0^{ij} \,,
$$
or
\be
 \xi_{\rm LS}(\theta,z_i,z_j) = \frac{1}{4\pi}\sum_{\ell =1}^{\ell_{\max}}(2\ell +1)C_\ell^{ij}P_\ell(\cos\theta)\,.
\ee
This is quite relevant as becomes clear when considering the radial and the transversal correlation function calculated in Ref.~\cite{Montanari:2012me}. There this monopole is not subtracted and it is found that the transversal correlation function is nearly entirely positive while the radial correlation function is nearly entirely negative.
Even though the integral of the theoretical correlation function over all of space, which is given by $P(0)$, vanishes for a Harrison-Zel'dovich spectrum, this is not the case, e.g., for the angular correlation function within a given a redshift slice or for the radial correlation function. It is therefore relevant whether the Landy--Szalay estimator for the correlation function is applied in each redshift slice or to the correlation function of the full survey.

Furthermore, even if $C_0^{ij}$ depends on quantities at the observer position (note that in Eq.~(\ref{Dez}) the not observable monopole and dipole terms which come from quantities at the observer position are already neglected), we can use the fact that these are equal for all redshifts $z_i, z_j$ and therefore differences like
$C^{ij}-C^{ii}$ are independent of the observer position. We consider especially
\be
\mathcal{M}(\bar z,\de z )\equiv \frac{1}{2} \sbra{C_0(z_-,z_-)+C_0(z_+,z_+)} - C_0(z_-,z_+)
\ee
where $z_\pm=\bar z\pm\de z/2$ (see \cite{Montanari:2012me}).
This quantity which is the angular average of the radial correlation function for galaxies at redshifts $z_+$ and $z_-$, contains interesting clustering information.

\begin{figure*}[tbp]
\centering
\includegraphics[width=0.4\textwidth]{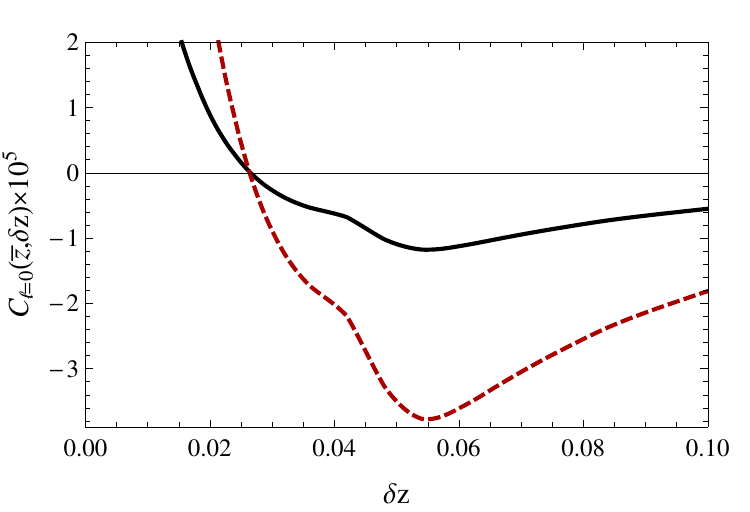}
\includegraphics[width=0.4\textwidth]{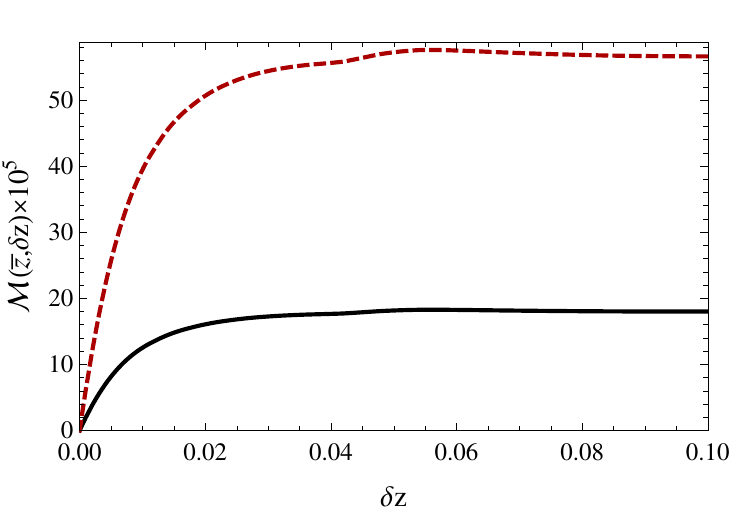} \\
\includegraphics[width=0.4\textwidth]{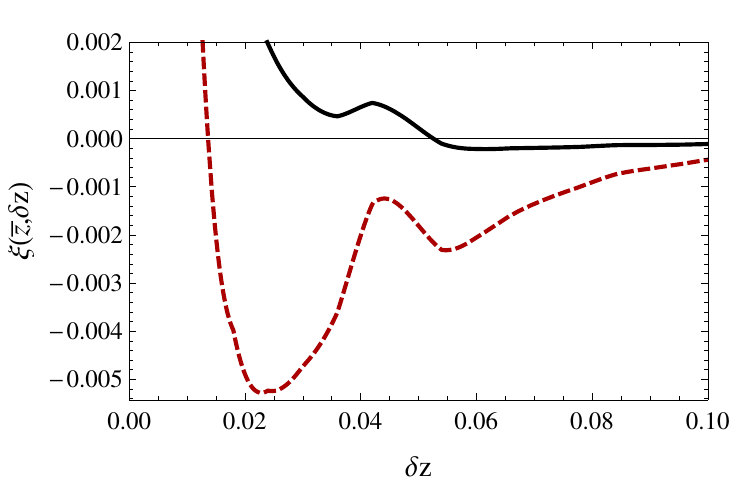}
\includegraphics[width=0.4\textwidth]{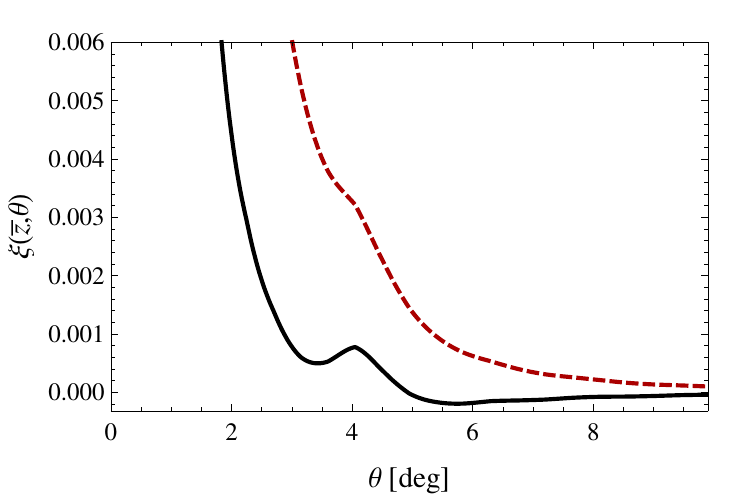}
\caption{
Top panel: the monopole $C_{\ell=0}(\bar z,\de z)$ and the observable monopole contribution $\mathcal{M}(\bar z,\de z)$ at mean redshift  $\bar z=0.55$.
Bottom panel: the radial (left) and the angular (right) correlation function at $\bar z=0.55$.
Solid (black) lines include only density correlations, dashed (red) lines also include redshift-space distortions.
The correlation functions are computed using Eq.~(\ref{e:corfun}); to wash out unphysical small scales oscillations, we use $\ell_s=80$ for $\xi(\bar z,\de z)$ and $\ell_s=400$ for $\xi(\bar z,\theta)$, as defined in Eq.~(2.20) of the accompanying paper~\cite{Didio1}.
}
\label{f:monopole}
\end{figure*}

In Fig.~\ref{f:monopole} we show the monopole as function of the radial redshift separation for a mean redshift $\bar z = 0.55$. Since here we are only interested in the theoretical modeling, we neglect redshift window functions and noise.
The observable monopole for $\de z=0$ vanishes by definition, and it tends to a constant for large $\de z$.

The difference between the density contribution only (black line) and full result (red dashed) is simply given by the redshift-space distortions, all other terms are very small for this case.
The Baryon Acoustic Oscillations (BAO) feature is clearly visible at $\de z\approx 0.045$ in $C_0(\bar z,\de z)$.
To confirm that this is indeed the BAO peak, the radial 2-point correlation function $\xi(\bar z,\de z)\equiv\xi(\theta=0,z_-,z_+)$ given by  Eq. (\ref{e:corfun}) but calculated from $\ell=0$ on (see \cite{Didio1} for more details) is also shown, presenting the same bump.
A detailed study of the BAO peak in the angular power spectrum is left for future work.
For completeness, also the transverse angular correlation function $\xi(\bar z,\theta)\equiv\xi(\theta,\bar z,\bar z)$ is plotted, which shows the BAO peak at $\theta\approx4^\circ$.
Contrary to the radial case in which the BAO scales have negative correlation, the transverse function is always positive on these scales.

Since we neglect shot noise and assume total sky coverage, cosmic variance from Eq.~(\ref{e:CosmVar}), $\sigma_{\ell=0}=\sqrt{2}C_{\ell=0}(\bar z,\de z)$ leads to a signal-to-noise
$$
\bra{\frac{S}{N}}_{\ell=0} = \frac{C_{\ell=0}(\bar z,\de z)}{ \sigma_{\ell=0}} = \frac{1}{\sqrt{2}} < 1,
$$
 for one given redshift difference $\de z$.
 This is also approximately the $(S/N)$ for $\MM$ for large enough $\de z$.
 Hence, if we add $\MM(\bar z,\de z)$ for several only weakly correlated redshifts we easily obtain a measurable signal with $S/N>1$.

\section{Conclusions}\label{s:con}
We have shown how the new code \class{gal}~\cite{Didio1} can be used to analyze galaxy surveys in an optimal way. With a few examples we have shown that the figure of merit from an analysis of the $C_\ell(z,z')$ spectra can be significantly larger, up to a factor of a few,  than the one from a standard $P(k)$ analysis.  This is due to the fact that this analysis makes optimal use of the redshift information and does not average over directions.
Clearly, in the analysis of upcoming high quality surveys we will want to use this promising method.

We have also seen that the nonlinearity scale,  the scale beyond which we can no longer trust the theoretically calculated spectrum, is of uttermost importance for the precision with which we estimate cosmological parameters. Within our conservative approach, and LSS data alone, cosmological parameters cannot be obtained with good precision, see Fig.~\ref{f:Ellipse_Euclid}. However, we hope that in the final data analysis we shall have accurate matter power spectra down to significantly smaller scales from N-body or approximation techniques. Furthermore, for an optimal determination of cosmological parameters, we shall of course combine LSS observations with CMB experiments, e.g.~from Planck, and other cosmological data.
As we have seen, the FoM of spectroscopic surveys increases significantly with the number of bins. However, the computational effort scales like $N_{\rm bin}^2$ and thus becomes correspondingly large. Nevertheless, the pre-factor in front of the scaling can be substantially smaller than one when we include only cross-correlations from bins within a given spatial distance.

We have also shown that deep angular galaxy catalogs can actually be used to measure the lensing potential. This is a novel method, alternative to the traditional lensing surveys, which can be used, e.g., to constrain modified gravity theories.

Since the power spectra of galaxy surveys depend on redshift, contrary to the CMB, here also the monopole contains cosmological information which can in principle be measured.

\acknowledgments{We thank Camille Bonvin,  Enrique Gazta\~naga, Ana\"\i s Rassat  and Alexandre Refregier for helpful discussions.
This work is supported by the Swiss National Science Foundation. RD acknowledges partial support from the (US) National Science Foundation under Grant No. NSF PHY11-25915. }

\appendix
\section{The galaxy number power spectrum}\label{appA}
The galaxy number counts in direction $\bf n$ at observed redshift $z$ are given by~\cite{Bonvin:2011bg,Didio1}

\bea
\De(\bn,z) &=& D +\left[\frac{1}{\HH}\dd_r(\bV\cdot\bn) + \left(\frac{\HH'}{\HH^2}+\frac{2}{r(z)\HH}\right)\bV\cdot\bn  -3\HH V \right]  \nonumber \\
&&+ \frac{1}{r(z)}\int_0^{r(z)}\hspace{-3mm}dr \left\{\left[2 - \frac{r(z)-r}{r}\Delta_\Om\right] \!(\Phi+\Psi)\right\}
\;  +\left(\frac{\HH'}{\HH^2}+\frac{2}{r(z)\HH} +1\right) \Psi \nonumber \\
&&+\frac{1}{\HH}\Phi'  -2 \Phi+ \left(\frac{\HH'}{\HH^2}+\frac{2}{r(z)\HH}\right)\left(\Psi+
 \int_0^{r(z)}\hspace{-3mm}dr(\Phi'+\Psi')\right) \,,
  \label{Dez}
\eea
where $V$ is the potential velocity defined through $\bV = - {\bf \nabla} V$.
Here primes denote derivatives w.r.t.~conformal time and
 the notation agrees with~\cite{Didio1}.
The first term is the density term, the term in square brackets is the redshift space distortion and the first term on the second line is the lensing term. The remaining gravitational potential contributions are sometimes also called ``relativistic terms''.

The $C_\ell$'s from this expression contain not only the auto-correlations of each term but also their cross-correlation with other contributions. We call the auto-correlation of the density term
$C^\de_\ell(z,z')$, the density term; the cross-correlation of density and velocity and the auto-correlation of the velocity term $C^z_\ell(z,z')$, the redshift space distortion term,   the cross-correlation of the lensing contribution with density and velocity and its  auto-correlation $C^{\rm lens}_\ell(z,z')$, the lensing term.  We call the rest the "potential terms", $C^{\rm pot}_\ell(z,z')$. If the cross-correlation terms dominate, any of these spectra except $C^\de_\ell$ can in principle be negative even for $z=z'$. These are the definitions of the parts of the full angular power spectra which are used in Section~\ref{s:lens}. More details on how these spectra are calculated can be found in the accompanying paper~\cite{Didio1}.

\section{Basics of Fisher matrix forecasts}\label{appB}

The Fisher matrix is defined as the derivative of the logarithm of the likelihood with respect to pairs of model parameters. Assuming that the spectra $C_\ell^{ij}$ are Gaussian (which is not a good assumption for small $\ell$ but becomes reasonable for $\ell \gtrsim 20$), the Fisher matrix is given by (cf.~\cite{Durrer:2008aa,Verde:2009tu})
\be
\label{eq:fisher}
F_{\alpha \beta} = \sum \frac{\partial C_\ell^{ij} }{\partial \lambda_\alpha} \frac{\partial C_\ell^{pq}}{\partial \lambda_\beta} \text{Cov}^{-1}_{\ell, (ij), (pq)},
\ee
where $\lambda_\alpha$ denotes the different cosmological parameters we want to constrain.
The sum over $\ell$ runs from 2 to a value $\ell_{\max}$ related to the non-linearity scale $k_{\max}$: we discuss this issue in section~\ref{ss:nonlin}.  Note also that we sum over the matrix indices $(ij)$ with $i\le j$ and $(pq)$ with $p\le q$ which run from 1 to
 $N_\text{bin}$ .

In the Fisher matrix approximation, i.e. assuming that the likelihood is a multivariate Gaussian with respect to cosmological parameters (which usually is not the case), the region in the full parameter space corresponding to a given Confidence Level (CL) is an ellipsoid centered on the best-fit model with parameters $\bar{\lambda}_\alpha$, with boundaries given by the equation $\sum_{\alpha,\beta} (\lambda_\alpha-\bar{\lambda}_\alpha)(\lambda_\beta-\bar{\lambda}_\beta)F_{\al\beta} = [\Delta \chi^2]$\footnote{The number $\Delta \chi^2$ depends both on the requested confidence level and on the number $n$ of parameters: for $n=1$ (resp. 2) and a 68\%CL one should use $\Delta \chi^2=1$ (resp. $2.3$). For other values see section 15.6 of~\cite{NR}.}, and with a volume given (up to a numerical factor) by $[\det(F^{-1})]^{1/2}$ (see e.g.~\cite{Durrer:2008aa}). Since the smallness of this volume is a measure of the performance of a given experiment, one often uses the inverse of the square root of the determinant as a Figure of merit,
$$\text{FoM} = \left[ \det  \left( F^{-1}\right) \right]^{-1/2}\,.$$
If we assume several parameters to be fixed by external measurements at the best-fit value $\bar{\lambda}_\alpha$, the 1$\sigma$ ellipsoid for the remaining parameters is given by the same equation, but with the sum running only over the remaining parameters. Hence the volume of this ellipsoid  is given by the square root of the determinant of the sub-matrix of $F$ restricted to the remaining parameters, that we call $\hat{F}$. In that case, the FoM for measuring the remaining parameters reads
$$\text{FoM}_{\text{fixed}} = \left[ \det  \left( (\hat{F})^{-1}\right) \right]^{-1/2}\,.$$
However, it is often relevant to evaluate how well one (or a few) parameters can be measured when the other parameters are marginalized over. In this case, a few lines of calculation show that the figure of merit is given by taking the sub-matrix of the inverse, instead of the inverse of the sub-matrix (see~\cite{Durrer:2008aa}),
$$\text{FoM}_{\text{marg.} } = \left[ \det  \left( \widehat{F^{-1}}\right) \right]^{-1/2}\,.$$
In particular, if we are interested in a single parameter $\la_\alpha$ and assume that all other parameters are marginalized over, the FoM for measuring $\la_\alpha$ is given by $$\text{FoM}_{\text{marg.} } = \left[(F^{-1})_{\alpha \alpha} \right]^{-1/2}\,.$$
When the likelihood is not a multivariate Gaussian with respect to cosmological parameters, the 68\% CL region is no longer an ellipsoid, but the FoM given above  (with the Fisher matrix being evaluated at the best-fit point) usually remains a good indicator.

It is however possible to construct examples where the FoM estimate completely fails. For instance, if two parameters are degenerate in a such a way that their profile likelihood is strongly non-elliptical (e.g.~with a thin and elongated banana shape). Then Fisher-based FoM will rely on a wrong estimate of the surface of the banana, and will return a very poor approximation of the true FoM. This happens e.g.~when including isocurvature modes~\cite{Ade:2013uln}, for mixed dark matter models~\cite{Boyarsky:2008xj} or in some modified gravity models~\cite{Audren:2013dwa}.

For the power spectrum analysis, following~\cite{Tegmark:1997rp,Seo:2003pu},  we define the Fisher matrix in each redshift bin as
\begin{equation} \label{FisherPk}
F_{\alpha \beta} \!\!=\!\! \int_{-1}^{1} \!\!\int_{k_{\min}}^{k_{\max}}\!\! \frac{\partial \ln P_{\text{obs}}}{\partial \lambda_\alpha} \frac{\partial \ln P_{\text{obs}}}{\partial \lambda_\beta} V_{\text{eff}} \frac{k^2 dk d\mu}{2 (2 \pi )^2} ,
\end{equation}
where the effective volume $V_{\text{eff}}$ is related to the actual volume $V_{\text{bin}}$ of each redshift bin through
\begin{equation}\label{FisherPkVol}
V_{\text{eff}} (k , \mu, \bar{z} ) = \left[ \frac{P_{\text{obs}}(k ,\mu, \bar{z} ) }{P_{\text{obs}}(k, \mu, \bar{z}) +1/\bar n(\bar{z}) } \right]^2 V_{\text{bin}}(\bar{z})\,.
\end{equation}
Here $\bar{z}$ is the mean redshift of the bin, and ${\bar n(\bar{z})}$ the average galaxy density in this bin, assumed to be uniform over the sky.
In the case of several non-overlapping $z$-bins, we assume that measurements inside each of them are independent so that the total Fisher matrix is the sum of those computed for every bin.
This expression is in principle valid in the flat-sky approximation. However, since it encodes all the statistical information, we can use it for a forecast analysis.
The denominator in Eq.~(\ref{FisherPkVol}) features the two contributions to the variance of the observable power spectrum  $P_{\text{obs}}(k ,\mu, \bar{z}  )$ coming from sampling variance and from shot noise.
The observable power spectrum $P_{\text{obs}}(k ,\mu, \bar{z}  )$ (not including shot noise) is given
in the minimal $\Lambda$CDM ($\La$ Cold Dark Matter) model by the
theoretical power spectrum $P(k)$ calculated at $z=0$, rescaled according to
\bea \label{e:P3}
P_{\text{obs}} (k ,\mu, \bar{z}  ) = \frac{\bar D_A (\bar{z})^2 H (\bar{z})}{D_A(\bar{z})^2 \bar H (\bar{z})} \left( 1 + \mu^2 \Omega_M(\bar{z})^{\gamma} \right)^2 G(\bar{z})^2 P(k)\,.
\eea
The first ratio in Eq.~(\ref{e:P3}) takes in account the volume difference for different cosmologies. The survey volume for galaxies in a redshift bin centered on $\bar{z}$ and of width $\de z$ is proportional to $D_A (\bar z)^2H(\bar z)^{-1}\de z$. The quantities $\bar D_A$ and $\bar H$ are evaluated at the fiducial cosmology.
The parenthesis contains the Kaiser approximation to redshift-space distortions \cite{Kaiser1987}, which together with the density term is the dominant contribution in Eq.~(\ref{Dez}) for our analysis. We assume an exponent $\gamma=0.6$ that is a good approximation to the growth factor from linear perturbation theory in  $\Lambda$CDM.
We have neglected the bias ($b=1$) in order to compare the results with the FoM derived from the angular power spectrum $C_\ell (z_1,z_2)$ where we also set $b=1$.
When we consider photometric redshift surveys, we need take into account the loss of information in the longitudinal direction due to  the redshift  error $\sigma_z$. Following~\cite{Seo:2003pu}, we then multiply $ P_\text{obs}(k, \mu,\bar z)$ with an exponential cutoff $e^{-\left( k \mu \sigma_z /H(z) \right)^2}$.

In Eq.~(\ref{FisherPk}), the observable power spectrum and the effective volume are assumed to be expressed in Hubble-rescaled units, e.g $[\textrm{Mpc}/h]^3$, while wave numbers are expressed in units of $[h/\textrm{Mpc}]$. In other words, it would be more rigorous to write everywhere ($[a_0^3 H_0^3  P_\textrm{obs}]$, $[k/(a_0 H_0)]$,  $[a_0^3 H_0^3 V_\mathrm{eff}]$) instead of ($P_\textrm{obs}$, $k$, $V_\mathrm{eff}$): the quantities in brackets are the dimensionless numbers that are actually measured, see~\cite{Lesgourgues:2013}.
Using Hubble-rescaled units does make a difference in the calculation of the partial derivative with respect to the model parameter $H_0$: it is important to keep $k/h$ and not $k$ constant. Also, the wavenumber $k_{\max}$ corresponding to the non-linearity scale is given in units $h/$Mpc,
so that we fix $k_{\max}/h$.

Furthermore, since the computation of $P(k)$ involves the assumption of a cosmological model to convert observable angles and redshifts into distances, expressing the latter in Mpc$/h$ mitigates the uncertainty introduced in this procedure since, to first approximation, distances $r(z)=\int dz/H(z)$ scale as $h^{-1}$.

\bibliography{CLASS-refs}
\bibliographystyle{JHEP}

\end{document}